\documentclass[12pt,letterpaper]{article}
\usepackage{graphicx}
\usepackage{accents}

\usepackage{lineno}
\usepackage{xcolor}

\def\title#1{{\bf\Large #1}}
\def\sec#1{\vspace{1.00\baselineskip} \noindent {\bf\large #1}
\vspace{0.25\baselineskip}}
\def\subsec#1{\vspace{0.50\baselineskip} \noindent {\bf #1}}

\def\ap{\approx}

\def\bt{\beta}
\def\ga{\gamma}

\def\de{\delta}

\def\ep{\epsilon}
\def\et{\eta}

\def\la{\lambda}
\def\La{\Lambda}

\def\om{\omega}
\def\Om{\Omega}
\def\pa{\parallel}
\def\pe{\perp}

\def\ph{\phi}

\def\Si{\Sigma}
\def\si{\sigma}

\def\ta{\tau}
\def\th{\theta}

\def\d{\dagger}
\def\<{\langle}
\def\>{\rangle}

\def\tsum{{\textstyle\sum}}

\def\d{\dagger}
\def\tr{{\rm tr}}
\def\diag{{\rm diag}}

\def\vk{\vec{k}}
\def\vn{\vec{n}}

\def\vr{\vec{r}}
\def\vu{\vec{u}}
\def\vv{\vec{v}}

\def\pa{\parallel}
\def\pe{\perp}
\def\ot{\accentset{\leftrightarrow}{\om}}
\def\st{\accentset{\leftrightarrow}{\si}}

\def\ba{\begin{eqnarray}}
\def\ea{\end{eqnarray}}
\def\be{\begin{equation}}
\def\ee{\end{equation}}

\begin{document}

%\centerline{Draft \hfill 2026/04/06 % 2023/07/29}

\begin{center}

\vspace*{0.0\baselineskip}

\title{Lorentz transformations in time and \\ two space dimensions: tutorial} %: \\ \vspace*{0.3\baselineskip} Matrix formalism}
%\title{Tutorial on Lorentz transformations in time \\ and two space dimensions}

\vspace{1.0\baselineskip}

C. J. McKinstrie \\

{\it\small Independent Photonics Consultant, Manalapan, NJ 07726, United States}

\vspace{0.50\baselineskip}

M. V. Kozlov \\

{\it\small Center for Preparatory Studies, Nazarbayev University, Astana 010000, Kazakhstan}

\vspace{0.50\baselineskip}

Abstract \\

\vspace{0.50\baselineskip}

\parbox[]{6.5in}{\small  In this article, matrix and vector formalisms for Lorentz transformations in time ($t$) and two space dimensions ($x$ and $y$) are developed and discussed.
Lorentz transformations conserve the squared interval $t^2 - x^2 - y^2$.
Examples of Lorentz transformations include boosts in arbitrary directions, which mix time and space, and rotations in space, which do not.
Lorentz transformations can be described by matrices and coordinate vectors. Lorentz matrices comprise the special orthogonal group SO(1,2).
The general form of a Lorentz matrix is derived, in terms of both components and block matrices.
Each Lorentz matrix $L$ has the Schmidt decomposition $QDP^t$, where $D$ is a diagonal matrix, and $P$ and $Q$ are orthogonal matrices.
It also has the Schmidt-like decomposition $R_2BR_1^t$, where $B$ is a boost matrix, and $R_1$ and $R_2$ are rotation matrices. Hence, a Lorentz matrix is specified by three parameters, namely the boost energy $\ga$, and the rotation angles $\th_1$ and $\th_2$.
Each Lorentz matrix has a pair of reciprocal Schmidt coefficients (which are real), and a unit coefficient, which is its own reciprocal.
It also has a pair of reciprocal eigenvalues (which are real or complex), and a unit eigenvalue.
The physical significances of the input and output Schmidt vectors, and the eigenvectors, are discussed.
Every Lorentz matrix (of physical interest) can be written as the exponential of a generating matrix. There are three basic generators, which produce boosts along the $x$ and $y$ axes, and a rotation about the $t$ axis (in the $xy$ plane).
These generators satisfy certain commutation relations, which show that SO(1,2) is homomorphic to Sp(2) and SU(1,1).
Simple formulas are derived for the energy and angle parameters of a composite transformation, in terms of the parameters of its constituent transformations.
This article is written at a level that is suitable for senior undergraduate students and junior graduate students.}

\end{center}

\newpage

\sec{1. Introduction}

The theory of special relativity was developed in the early 1900s \cite{lor04,ein05,ein23} and remains a fascinating subject.
Undergraduate students are taught about relativity by the use of simple one-space-dimensional examples \cite{tay05,tip08}. Particle collisions are analyzed in two dimensions, but one can always Lorentz transform to a frame in which the collisions are one-dimensional, in the (parallel) direction of approach, and the individual momentum components in the other (perpendicular) direction are conserved. A similar statement can be made about Doppler shifts. Graduate students are taught the formalism for Lorentz transformations in time and three space dimensions, which involves covariant and contravariant vectors, and the metric tensor \cite{lan75,jac99,gol02}. Although this formalism is four-dimensional (in time and space), it remains true that most of the examples used to illustrate it are one-space-dimensional. Complexity manifests itself when successive transformations are considered: In one dimension, every transformation is a boost, transformations are always in the same (or opposite) direction and their composition rules are simple. However, in higher dimensions, transformations are combinations of boosts and rotations, and the boosts can act in different directions. In two dimensions, a third boost is coplanar with the first two, whereas in three dimensions, the third boost can be noncoplanar. The complexity of the mathematics obscures the underlying physics. In this article, we review Lorentz transformations in time and two space dimensions, because the mathematical complexity is intermediate betweeen the complexities of one- and three-dimensional transformations, and the physics is interesting.

This article is organized as follows:
To introduce the basic concepts of Lorentz transformations, in Sec. 2 we discuss transformations in time and one space dimension. These transformations can be written in the matrix form $T' = LT$, where $L$ is a $2 \times 2$ matrix, and $T = [t, x]^t$ and $T' = [t', x']^t$ are $2 \times 1$ input and output coordinate vectors, respectively. (Time and distance are measured in the same units, so time is really $ct$ or distance is really $x/c$, and velocity is really $v/c$.) The Lorentz matrix $L$ is specified by two (dimensionless) parameters, the velocity $\bt$ and energy $\ga = 1/(1 - \bt^2)^{1/2}$, only one of which is independent (free). Lorentz transformations preserve the squared interval $(t')^2 - (x')^2 = t^2 - x^2$.
Standard formulas, for the addition and subtraction of velocities, are derived and discussed briefly.
The set of Lorentz matrices forms a group under multiplication, which is called the special orthogonal group SO(1,1). Every Lorentz matrix (of practical interest) can be written as the exponential of a generating matrix: $L = \exp(Gk)$, where $k$ is a free parameter [$\bt = \tanh(k)$].

In two space dimensions, Lorentz transformations are specified by $3 \times 3$ matrices and coordinates are specified by $3 \times 1$ vectors. The Lorentz condition is $L^tSL = S$, where $S = \diag(1,-1,-1)$ is the metric (structure) matrix. This condition ensures that the squared interval $T^tST = t^2 - x^2 - y^2$ is conserved. In Secs. 3.1 and 3.2, we use the Lorentz condition to show that Lorentz matrices form a group, which is called the special orthogonal group SO(1,2). Matrices that describe boosts in the $x$ and $y$ directions, and rotations about the $t$ axis (in the $xy$ plane) are members of this group.

In Secs. 3.3 and 3.4, we discuss combinations of boosts and rotations. Although the combination of two rotations is another rotation, and the combination of a boost and two inverse rotations is a boost in a different direction, the combination of two boosts is not necessarily another boost.

In Sec. 3.5, we use the Lorentz condition to derive the general form of a Lorentz matrix, which is a boost matrix followed, or preceded, by a rotation matrix \cite{wig39}. The boost matrix is specified by two free parameters, namely the boost energy (or momentum) and direction (angle), whereas the rotation matrix is specified by one free parameter, the rotation angle.

Every real matrix $M$ has the singular-value (Schmidt) decomposition $M = QDP^t$, where $D$ is a diagonal matrix, and $P$ and $Q$ are orthogonal matrices \cite{hor13}. The components of $D$ are called Schmidt coefficients, and the columns of $P$ and $Q$ are called input and output Schmidt vectors, respectively.
In Sec. 3.6, we use the general form of a Lorentz matrix to determine its Schmidt decomposition. The coefficients, and the associated input and output vectors, all have clear physical significances, which elucidate the transformation.

In Sec. 3.7, we use a Schmidt-like decomposition of a Lorentz matrix to determine its spectral decomposition. Real eigenvalues correspond to dilations in the directions of the associated real eigenvectors, whereas a pair of complex conjugate eigenvalues corresponds to rotation about an axis that is specified by the real and imaginary parts of the associated complex eigenvectors.

In Sec. 3.8, we show that every Lorentz matrix $L$ (of practical interest) can be written as the exponential of a generating matrix $G$, which is a linear combination of three basis generators \cite{sat86,gil08}. These generators represent boosts in the $x$ and $y$ directions, and a rotation. Thus, an arbitrary transformation can be built from boosts and rotations, which are easy to understand.
We derive formulas that relate the boost and angle parameters to the generator coefficients (parameters).

In Sec. 3.9, we consider the combination of two arbitrary transformations. Formulas are derived, which specify the parameters of the combined transformation in terms of the parameters of the composite transformations.

The main results of this article are summarized in Sec. 4.
Finally, for convenience, the scalar Lorentz-transformation formulas are derived from first principles in App. A, and the vector formulas are derived and discussed in Apps. B and C.

\newpage

\sec{2. Lorentz transformations in one dimension}

In this section, we discuss Lorentz transformations in time and one space dimension, to introduce important physical concepts in simple mathematical terms.
The theory of special relativity is based on the axiom that the laws of physics are the same in any inertial frame \cite{ein05,ein23,tay05,tip08}. No frame is privileged.
Consider transformations among three frames, namely the laboratory frame (LF0) and two moving frames (MF1 and MF2).
Suppose that the axes of the three frames are aligned, and MF1 and MF2 move along the common $x$-axis.
In one dimension, the coordinates of a point are $t$ and $x$. The forward transformation between LF0 and MF1 is specified by the equations
\be t_1 = \ga_{10}(t_0 - \bt_{10}x_0), \ \ x_1 = \ga_{10}(x_0 - \bt_{10}t_0), \label{2.1} \ee
where $\bt_{10}$ is the velocity of MF1 relative to LF0 (in which it is measured), and the energy $\ga_{10} = 1/(1 - \bt_{10}^2)^{1/2}$. In these equations, time and space have the same units (so time is really $ct$ or distance is really $x/c$), and the velocity and energy are dimensionless (so velocity is really $v/c$). The squared interval $\Si = t^2 - x^2$. It follows from Eqs. (\ref{2.1}) that $\Si_1 = t_1^2 - x_1^2 = t_0^2 - x_0^2 = \Si_0$. Thus, transformation (\ref{2.1}) preserves the squared interval, as it was designed to do (App. A).
By inverting Eqs. (\ref{2.1}), one obtains the backward transformation equations
\be t_0 = \ga_{10}(t_1 + \bt_{10}x_1), \ \ x_0 = \ga_{10}(x_1 + \bt_{10}t_1). \label{2.2} \ee
One can rewrite Eqs. (\ref{2.2}) in the forms of Eqs. (\ref{2.1}) by defining $\bt_{01} = -\bt_{10}$. Observers in the two frames agree on the magnitude of their relative velocity, but disagree on its sign. (If MF1 appears to move to the right relative to LF0, then LF0 appears to move to the left relative to MF1.)

For reference, the energy depends on the squared velocity, which is non-negative, and the momentum $u_{10}$ is the product of the energy and velocity. Hence, $\ga_{01} = \ga_{10}$ and $u_{01} = -u_{10}$. The observers also agree on the magnitude of their relative momentum. Written in terms of energy and momentum (instead of velocity), the forward transformation equations are
\be t_1 = \ga_{10}t_0 - u_{10}x_0, \ \ x_1 = \ga_{10}x_0 - u_{10}t_0, \label{2.3} \ee
and the backward transformation equations are
\be t_0 = \ga_{10}t_1 + u_{10}x_1, \ \ x_0 = \ga_{10}x_1 + u_{10}t_1, \label{2.4} \ee
where $\ga_{10} = (1 + u_{10}^2)^{1/2}$.
The transformations between LF0 and MF2, and MF1 and MF2, are similar.

One can use Eqs. (\ref{2.1}) and (\ref{2.2}) to derive the velocity addition and subtraction formulas.
Suppose that an object is at rest in MF2 (at the origin). Then, in LF0 it moves at velocity $\bt_{20}$ and its incremental coordinate vector, which represents a small displacement in a short time, is $(dt_0, \bt_{20}dt_0)$. It follows from Eqs. (\ref{2.1}) that
\ba dt_1 = \ga_{10}(dt_0 - \bt_{10}\bt_{20}dt_0), \ \ dx_1 = \ga_{10}(\bt_{20}dt_0 - \bt_{10}dt_0). \label{2.5} \ea
By dividing the right sides of Eqs. (\ref{2.5}), one finds that
\be \bt_{21} = {dx_1 \over dt_1} = {\bt_{20} - \bt_{10} \over 1 - \bt_{10}\bt_{20}}. \label{2.6} \ee
Equation (\ref{2.6}) is called the velocity subtraction (or relative velocity) formula. If MF1 moves in the same direction as the object, then the object's speed in MF1 is lower than its speed in LF0.
Alternatively, if the incremental coordinate vector in MF1 is $(dt_1,\bt_{21}dt_1)$, then it follows from Eqs. (\ref{2.2}) that
\be dt_0 = \ga_{10}(dt_1 + \bt_{10}\bt_{21}dt_1), \ \ dx_0 = \ga_{10}(\bt_{21}dt_1 + \bt_{10}dt_1). \label{2.7} \ee
By dividing the right sides of Eqs. (\ref{2.7}), one finds that
\be \bt_{20} = {dx_0 \over dt_0} = {\bt_{10} + \bt_{21} \over 1 + \bt_{10}\bt_{21}}. \label{2.8} \ee
Equation (\ref{2.8}) is called the velocity addition formula.
If MF1 moves in the same direction as an object, then the object's speed in LF0 is higher than its speed in MF1.
Notice that in the addition and subtraction formulas, $\bt_{10}$ and $\bt_{20}$ are measured in LF0, whereas $\bt_{21}$ is measured in MF1.

Equations (\ref{2.6}) and (\ref{2.8}) are consistent, in the sense that if one solves Eq. (\ref{2.6}) for $\bt_{20}$, one obtains formula (\ref{2.8}), and if one solves Eq. (\ref{2.8}) for $\bt_{21}$, one obtains formula (\ref{2.6}).
One can deduce the latter equation from the former, or the former equation from the latter, by interchanging the subscripts 0 and 1, and using the fact that $\bt_{01} = -\bt_{10}$. This result is consistent with the relativistic equivalence of LF0 and MF1.

In two-dimensional geometry, there are two kinds of rotation. In a passive rotation, the position vector $\vr$ stays the same and the coordinate system used to represent it rotates through the angle $-\th$, whereas in an active rotation, the coordinate system stays the same and the position vector rotates through the angle $\th$. Similar statements can be made about Lorentz transformations. Equations (\ref{2.3}) and (\ref{2.4}) describe passive transformations, because the space-time point does not change, but the coordinate system (frame) used to represent it does. Consider Eqs. (\ref{2.4}). In the passive picture, LF0 moves to the left relative to MF1, so a particle that is at rest in MF1 appears to move to the right, with momentum $u_{10}$, in LF0. In the active picture, there is only one frame, LF0, and the transformation increases (boosts) the particle momentum from 0 to $u_{10}$. Henceforth, we will consider mainly active transformations, because they are easier to visualize. In this picture, the coordinates specify the space-time state.

It is convenient to use matrices to model Lorentz transformations. Equations (\ref{2.4}) can be rewritten the matrix form
\be T_1 = L_{10}T_0, \label{2.11} \ee
where $T_0 = [t_0, x_0]^t$ and $T_1 = [t_1, x_1]^t$ are $2 \times 1$ coordinate vectors, and
\be L_{10} = \left[\begin{array}{cc} \ga_{10} & u_{10} \\ u_{10} & \ga_{10} \end{array}\right] \label{2.12} \ee
is the $2 \times 2$ transformation matrix, whose components satisfy the auxiliary equation $\ga_{10}^2 - u_{10}^2 = 1$.
What kind of transformation does Eq. (\ref{2.11}) represent?
If an object is stationary in LF0, then its differential coordinate vector $dT_0 = [dt_0, 0]^t$. In MF1, $dT_1 = [\ga_{10}dt_0, u_{10}dt_0]^t$, so its velocity $dx_1/dt_1 = u_{10}/\ga_{10} = \bt_{10}$. Thus, Eq. (\ref{2.11}) describes to a boost, as stated above.
Not only does Eq. (\ref{2.11}) apply to coordinate vectors, it also applies to other two-vectors. For example, let $E_0 = [1, 0]^t$ be the energy-momentum two-vector of a particle of unit mass, which is initially at rest. Then after the transformation, $E_1 = [\ga_{10}, u_{10}]^t$, which is the first column of $L_{10}$. The final particle momentum equals the momentum used to define the transformation.

The Lorentz matrix (\ref{2.12}) has several important properties. First, $\det(L_{10}) = \ga_{10}^2 - u_{10}^2$ $= 1$ and second, $L_{10}^{-1}(u_{10}) = L_{10}(-u_{10})$. The inverse of a boost is a brake of the same magnitude. Let $S = \diag(1,-1)$ be the metric (structure) matrix. Then the squared interval $\Si = T^tST$. It is easy to verify that $L_{10}^tSL_{10} = S$, from which it follows that $T_1^tST_1 = T_0^tL_{10}^tSL_{10}T_0 = T_0^tST_0$. Matrix (\ref{2.12}) conserves the squared interval, as it should do. It is also easy to verify that $SL_{10}^t(u_{10})S = L_{10}(-u_{10}) = L^{-1}(u_{10})$.
In one dimension, Lorentz matrices are symmetric. We included the transpose symbol in the preceding equations because the matrices that describe higher-dimensional transformations are not necessarily symmetric (Sec. 3).
The Lorentz matrices $L_{20}$ and $L_{21}$ are defined in similar ways, and have similar properties.

If one were to transform state 0 to state 2 directly, one would write $T_2 = L_{20}T_0$. Alternatively, if one were to transform state 0 to state 2 via state 1, one would write $T_2 = L_{21}T_1 = L_{21}L_{10}T_0$. Hence,
\be L_{20} = L_{21}L_{10} = \left[\begin{array}{cc} \ga_{21}\ga_{10} + u_{21}u_{10} & \ga_{21}u_{10} + u_{21}\ga_{10} \\ u_{21}\ga_{10} + \ga_{21}u_{10} & \ga_{21}\ga_{10} + u_{21}u_{10} \end{array}\right]. \label{2.13} \ee
One obtains the matrix for a composite transformation by multiplying the matrices for the constituent transformations.
Notice that $\det(L_{20}) = \det(L_{21})\det(L_{10}) = 1$, so the product matrix has same form as the consituent matrices, in which the diagonal and off-diagonal components represent energy and momentum, respectively, and satisfy the aforementioned auxiliary equation.
If the initial (incremental) state vector $dT_0 = [dt_0, 0]^t$, then the final state vector $dT_2 = [(\ga_{21}\ga_{10} + u_{21}u_{10})dt_0, (u_{21}\ga_{10} + \ga_{21}u_{10})dt_0]^t$, from which it follows that
\be \bt_{20} = {dx_2 \over dt_2} = {\bt_{21} + \bt_{10} \over 1 + \bt_{21}\bt_{10}}. \label{2.14} \ee
The ratio in formula (\ref{2.14}) is the bottom-left component of $L_{20}$ divided by the top-left component ($E_2 = L_{20}E_0$). Formula (\ref{2.14}) represents two successive boosts and is consistent with Eq. (\ref{2.8}).

By multiplying the equation $L_{20} = L_{21}L_{10}$ by $L_{10}^{-1}$ on the right, one finds that
\be L_{21} = L_{20}L_{10}^{-1} = \left[\begin{array}{cc} \ga_{20}\ga_{10} - u_{20}u_{10} & u_{20}\ga_{10} - \ga_{20}u_{10} \\ u_{20}\ga_{10} -\ga_{20}u_{10} & \ga_{20}\ga_{10} - u_{20}u_{10} \end{array}\right]. \label{2.15} \ee
If the initial (incremental) state vector $dT_1 = [dt_1, 0]^t$, then the final state vector $dT_2 = [(\ga_{21}\ga_{10} - u_{21}u_{10})dt_1, (u_{21}\ga_{10} - \ga_{21}u_{10})dt_1]^t$, from which it follows that
\be \bt_{21} = {dx_2 \over dt_2} = {\bt_{20} - \bt_{10} \over 1 - \bt_{20}\bt_{10}}. \label{2.16} \ee
The ratio in formula (\ref{2.16}) is the bottom-left component of $L_{21}$ divided by the top-left component ($E_2 = L_{21}E_1$). Formula (\ref{2.16}) represents a brake followed by a boost and is consistent with Eq. (\ref{2.6}). Notice that $L_{21}L_{10} = L_{10}L_{21}$ and $L_{20}L_{10}^{-1} = L_{10}^{-1}L_{20}$, so in one dimension, the transformation matrices commute.

In the preceding analyses, every Lorentz matrix $L(u)$ is specified by one free parameter, the momentum. The identity matrix $I = L(0)$ and the inverse matrix $L^{-1}(u) = L(-u)$ are Lorentz matrices. If $L_{10}$ and $L_{21}$ are Lorentz matrices, then so also is the product matrix $L_{21}L_{10}$. Matrix multiplication is associative, so $L_{32}(L_{21}L_{10}) = (L_{32}L_{21})L_{10} = L_{32}L_{21}L_{10}$. Hence, the set of Lorentz matrices forms a group under multiplication \cite{taw95}, which is called the special orthogonal group SO(1,1). This group is simple and needs no further discussion.

Now consider the matrix
\be G = \left[\begin{array}{cc} 0 & 1 \\ 1 & 0 \end{array}\right], \label{2.21} \ee
which has the property $G^2 = I$. It is easy to verify that
\ba \exp(Gk) &= &I + Gk + Ik^2/2 + Gk^3/3! \dots \nonumber \\
&= &I\cosh(k) + G\sinh(k), \label{2.22} \ea
where $k$ is a free parameter. Written in terms of components,
\be \exp(Gk) = \left[\begin{array}{cc} C & S \\ S & C \end{array}\right], \label{2.23} \ee
where $C = \cosh(k)$, $S = \sinh(k)$ and $C^2 - S^2 = 1$. Matrix (\ref{2.23}) has the form of matrix (\ref{2.12}), with $\ga = C$ and $u = S$. Thus, every Lorentz matrix can be written as the exponential of matrix (\ref{2.21}). For this reason, $G$ is called the generating matrix (generator) and $k$ is called the generator coefficient. There is a simple relation between this coefficient and the momentum. It is easy to verify that
\be \exp(Gk_2)\exp(Gk_1) = \exp[G(k_2 + k_1)]. \label{2.24} \ee
In terms of generators, the product rule is simple: The coefficient of the product matrix is the sum of the coefficients of the constituent matrices. [Compare this result to Eq. (\ref{2.13}).]

To summarize the main points of this section, Lorentz transformations [Eqs. (\ref{2.11}) and (\ref{2.12})] conserve the squared interval. Every Lorentz matrix can be written as the exponential of a generating matrix. The transformation matrix associated with a composite transformation is the product of the matrices associated with the constituent transformations. Matrix algebra is well known, so its use facilitates the study of composite transformations. In one dimension, the composition of two boosts is another boost and the order of the boosts does not matter. These statements are not true in higher dimensions (Sec. 3).

In this section, we used the precise notation $T_f = L_{fi}(u_{fi})T_i$, where the subscripts $i$ and $f$ denote the initial and final states (or frames), respectively, and $u_{if} = -u_{fi}$. In the rest of this article, we use the concise notation $T_n = L_n(u_n)T_{n-1}$, in which the subscript $n$ is the number of the transformation, or $T' = L(u)T$, where the prime $'$ denotes the output state.

\newpage

\sec{3. Lorentz transformations in two dimensions}

In this section, we develop a matrix formalism for Lorentz transformations in time ($t$) and two space dimensions ($x$ and $y$). In Sec. 3.1, we discuss the basic properties of these transformations, which are determined by the requirement that they conserve the squared interval $\Si = t^2 - x^2 - y^2$. In Sec. 3.2, we discuss two examples of Lorentz transformations, namely boosts and rotations. In Sec. 3.3, we discuss combinations of boosts and rotations, whereas in Sec. 3.4, we discuss combinations of boosts.
In Sec. 3.5, we derive the general form of a Lorentz matrix, which shows that an arbitrary transformation consists of a boost followed by a rotation (or the same rotation followed by a different boost). In Secs. 3.6 and 3.7, we use this general form to determine the Schmidt and spectral decompositions of an arbitrary Lorentz matrix, respectively. In Sec. 3.8, we show that every Lorentz matrix $L$ (of practical interest) can be written as the exponential of a generating matrix $G$. Finally, in Sec. 3.9, we discuss combinations of arbitrary transformations.

\subsec{3.1 Basic properties of transformations}

Let $L$ be a real $3 \times 3$ matrix, $T = [t, x, y]^t$ be a $3 \times 1$ coordinate vector and $S = \diag(1, -1, -1)$ be the metric (structure) matrix. Then the squared interval $\Si = T^tST$. If light is emitted by a source at the origin ($x$, $y = 0$ at $t = 0$), then its leading edge expands radially outward at unit speed. The radius of the wavefront $r = t$, which is the equation for a cone, and which is equivalent to the equation $\Si = 0$. (For this reason, $\Si$ is also called the light-cone variable.)

A Lorentz transformation from one coordinate system to another is specified by the equation $T' = LT$, where $L$ is the transformation matrix. In the primed system, $\Si' = (T')^tST' = TL^tSLT$. Hence, the squared interval is conserved (which is a fundamental tenet of special relativity) if and only if
\be L^tSL = S. \label{3.1.1} \ee
Matrices that satisfy Eq. (\ref{3.1.1}) are called Lorentz matrices.
Trivial examples of such matrices include the identity matrix $I$ and $S$ itself (because $S^t = S$ and $S^2 = I$). If Eq. (\ref{3.1.1}) is satisfied, then $S(L^tSL) = S^2 = I$. Hence, $L$ has the inverse
\be L^{-1} = SL^tS. \label{3.1.2} \ee
Conversely, if Eq. (\ref{3.1.2}) is satisfied, then $S = SL^{-1}L = L^tSL$. Hence, Eqs. (\ref{3.1.1}) and (\ref{3.1.2}) are equivalent. When one tries to establish the properties of Lorentz matrices, one can use whichever equation is more convenient.

For example, if $L$ is Lorentzian, %is a Lorentz matrix (is Lorentzian),
then $(L^{-1})^{-1} = S^{-1}(L^t)^{-1}S^{-1} = S(L^{-1})^tS$. In addition, $S = LL^{-1}S = LSL^tS^2 = LSL^t$. Hence, $L^{-1}$ and $L^t$ are also Lorentzian.
If $L_1$ and $L_2$ are both Lorentzian, then $(L_2L_1)^tS(L_2L_1) = L_1^t(L_2^tSL_2)L_1 = L_1^tSL_1 = S$. Hence, $L_2L_1$ is also Lorentzian. By using the principle of induction, one can extend this result to the product of an arbitrary number of Lorentz matrices. It follows from the laws of matrix multiplication that $L_3(L_2L_1) = (L_3L_2)L_1 = L_3L_2L_1$.

To summarize these properties, the set of Lorentz matrices is closed under multiplication, which is associative. The set contains the identity matrix, and every member of the set has an inverse. Hence, the set of Lorentz matrices is a group under multiplication \cite{taw95}, which is called the orthogonal group O(1,2). The metric matrix $S$ has an indefinite signature (one positive eigenvalue and two negative eigenvalues) and its presence in the Lorentz condition (\ref{3.1.1}) imposes constraints on $L$. For this reason, O(1,2) is also called the indefinite orthogonal group. $S$ is also the metric matrix for the associated vector space \cite{taw91}, whose members are the coordinate vectors $T$. Because $S$ is indefinite, the generalized inner product (squared interval) $T^tST$ can be positive, zero or negative.

Three further comments are in order.
First, the Lorentz condition (\ref{3.1.1}) is symmetric [$(L^tSL)^t$ $= L^tSL$], so it imposes six constraints on the nine components of $L$. Hence, the members of O(1,2) are specified by (up to) three independent (free) parameters.
Second, it follows from the Lorentz condition and the determinant rule that $\det(L^tSL) = \det(L^t)\det(S)\det(L) = \det(S)$. Hence,  $[\det(L)]^2 = 1$. The set of matrices with determinant $-1$ is not a group, because it is not closed under multiplication. We consider only the set of matrices with determinant 1, which is closed under multiplication. This subgroup is called the special orthogonal group SO(1,2).
Third, let $L = [l_{ij}]$, where the indices $i$ and $j$ vary from 0 to 2. Then it also follows from the Lorentz condition that $l_{00}^2 - l_{10}^2 - l_{20}^2 = 1$, which requires that $l_{00} \ge 1$ or $l_{00} \le -1$ \cite{sat86}. The set of matrices for which $l_{00} \ge 1$ is a subgroup of SO(1,2), which is called the restricted Lorentz group. Its members describe physical transformations. Henceforth, we will omit the term restricted.

\newpage

\subsec{3.2 Examples of transformations}

Nontrivial examples of Lorentz matrices include the matrices
\be L_x(u) = \left[\begin{array}{ccc} \ga & u & 0 \\ u & \ga & 0 \\ 0 & 0 & 1 \end{array}\right], \ \ 
L_y(u) = \left[\begin{array}{ccc} \ga & 0 & u \\ 0 & 1 & 0 \\ u & 0 & \ga \end{array}\right], \label{3.2.1}\ee
where $u$ and $\ga = (1 + u^2)^{1/2}$ are the (dimensionless) momentum and energy parameters, respectively.
(Notice that $u$ can be positive, zero or negative.)
It is easy to verify that $\det(L_x) = \det(L_y) = 1$. Not only do matrices (\ref{3.2.1}) transform the coordinate three-vector $T$, they also transform $E_p = [\ga_p, u_{px}, u_{py}]^t$, which is the energy-momentum three-vector of a particle of unit mass. Suppose that the particle is at rest, in which case $E_p = [1, 0, 0]^t$. Then after the transformation $E_p' = L_x(u)E_p$, the three-vector
\be \left[\begin{array}{c} \ga_p' \\ u_{px}' \\ u_{py}' \end{array}\right]
= \left[\begin{array}{ccc} \ga & u & 0 \\ u & \ga & 0 \\ 0 & 0 & 1 \end{array}\right]
\left[\begin{array}{c} 1 \\ 0 \\ 0 \end{array}\right]
= \left[\begin{array}{c} \ga \\ u \\ 0 \end{array}\right]. \label{3.2.2} \ee
The particle has momentum $u$ in the $x$ direction. In a similar way, the transformation $E_p' = L_y(u)E_p$ produces a particle with momentum $u$ in the $y$ direction. The matrices in Eq. (\ref{3.2.1}) represent active transformations (boosts). Henceforth, we will use the symbol $B$ to denote boost matrices and reserve $L$ for general transformation matrices (the natures of which remain to be determined).
It follows from a standard matrix formula that $B^{-1}(u) = B(-u)$. The inverse of a boost is a brake of the same magnitude (a boost in the opposite direction). This result makes physical sense. Notice that the boost matrices $B_x$ and $B_y$ are symmetric.

Another nontrivial example is the rotation matrix
\be R(\th) = \left[\begin{array}{ccc} 1 & 0 & 0 \\ 0 & c & -s \\ 0 & s & c \end{array}\right], \label{3.2.3} \ee
where $c = \cos\th$, $s = \sin\th$ and $\th$ is the rotation angle. It is also easy to verify that $\det(R) = 1$. Rotations do not affect the time components of coordinate three-vectors. In shortened ($2 \times 2$) notation, let $[1, 0]^t$ and $[0, 1]^t$ be unit vectors in the $x$- and $y$-directions, respectively. Then the rotated vectors
\be \left[\begin{array}{cc} c & -s \\  s & c \end{array}\right] \left[\begin{array}{c} 1 \\ 0 \end{array}\right]
= \left[\begin{array}{c} c \\ s \end{array}\right], \ \ 
\left[\begin{array}{cc} c & -s \\  s & c \end{array}\right] \left[\begin{array}{c} 0 \\ 1 \end{array}\right]
= \left[\begin{array}{c} -s \\ c \end{array}\right]. \label{3.2.4} \ee
The matrix in Eq. (\ref{3.2.3}) represents an active (positive) rotation through the angle $\th$.
It follows from the same matrix formula that $R^{-1}(\th) = R(-\th) = R^t(\th)$. The opposite of a positive rotation is a negative rotation  through the same angle. This result also makes physical sense. Notice that rotation matrices are asymmetric and orthogonal ($R^{-1} = R^t$).

There are two additional examples, which clarify some results in the following sections. Let $E = [\ga, uc_1, us_1]^t$ be the energy-momentum three-vector associated with the direction $(c_1, s_1)$, where $c_1 = \cos\th_1$ and $s_1 = \sin\th_1$. Then
\be \left[\begin{array}{ccc} 1 & 0 & 0 \\ 0 & c_2 & -s_2 \\ 0 & s_2 & c_2 \end{array}\right]
\left[\begin{array}{c} \ga \\ uc_1 \\ us_1 \end{array}\right]
= \left[\begin{array}{c} \ga \\ u(c_2c_1 - s_2s_1) \\ u(s_2c_1 + c_2s_1) \end{array}\right]. \label{3.2.5} \ee
The rotation matrix, acting to the right, rotates the spatial part of the column three-vector (the momentum vector) by $\th_2$ radians. Furthermore,
\be \left[\begin{array}{ccc} \ga & uc_1 & us_1 \end{array}\right]
\left[\begin{array}{ccc} 1 & 0 & 0 \\ 0 & c_2 & -s_2 \\ 0 & s_2 & c_2 \end{array}\right]
= \left[\begin{array}{ccc} \ga & u(c_1c_2 + s_1s_2) & u(s_1c_2 - c_1s_2) \end{array}\right]. \label{3.2.6} \ee
The rotation matrix, acting to the left, rotates the spatial part of the row three-vector by $-\th_2$ radians. This inverse rotation is a consequence of the equation $(C^tR)^t = R^tC = R^{-1}C$, where $C$ is a column vector.

\newpage

\subsec{3.3 Combinations of boosts and rotations}

Let $R(\th_1)$ and $R(\th_2)$ be matrices of the form (\ref{3.2.3}), where $\th_i$ is a rotation angle. Then their product
\be \left[\begin{array}{ccc} 1 & 0 & 0 \\ 0 & c_2 & -s_2 \\ 0 & s_2 & c_2 \end{array}\right]
\left[\begin{array}{ccc} 1 & 0 & 0 \\ 0 & c_1 & -s_1 \\ 0 & s_1 & c_1 \end{array}\right]
= \left[\begin{array}{ccc} 1 & 0 & 0 \\ 0 & c_2c_1 - s_2s_1 & -s_2c_1 - c_2s_1 \\ 0 & s_2c_1 + c_2s_1 & c_2c_1 - s_2s_1 \end{array}\right]. \label{3.3.1} \ee
Hence, $R(\th_2)R(\th_1) = R(\th_2 + \th_1)$. Notice that rotation matrices commute (in two space dimensions). Notice also that $R(\th_2)R^{-1}(\th_1) = R(\th_2 - \th_1)$.

The most important combination of boost and rotation matrices arises in the context of a boost in an arbitrary direction. How does one specify such a boost? First, one rotates the coordinate system by $\th$ radians, so that the $x$-axis is parallel to the boost direction. Second, one applies a boost of magnitude $u$. Third, one rotates the coordinate system back to its original orientation. Hence, $B(u,\th) = R^{-1}(\th)B(u)R(\th)$, which is a similarity transform of $B(u)$. Written explicitly,
\ba B(u,\th) &= &\left[\begin{array}{ccc} 1 & 0 & 0 \\ 0 & c & -s \\ 0 & s & c \end{array}\right]
\left[\begin{array}{ccc} \ga & u & 0 \\ u & \ga & 0 \\ 0 & 0 & 1 \end{array}\right]
\left[\begin{array}{ccc} 1 & 0 & 0 \\ 0 & c & s \\ 0 & -s & c \end{array}\right] \nonumber \\
&= &\left[\begin{array}{ccc} 1 & 0 & 0 \\ 0 & c & -s \\ 0 & s & c \end{array}\right]
\left[\begin{array}{ccc} \ga & uc & us \\ u & \ga c & \ga s \\ 0 & -s & c \end{array}\right] \nonumber \\
&= &\left[\begin{array}{ccc} \ga & uc & us \\ uc & \ga c^2 + s^2 & \de cs \\ us & \de sc & \ga s^2 + c^2 \end{array}\right] \nonumber \\
&= &\left[\begin{array}{ccc} \ga & uc & us \\ uc & 1 + \de c^2 & \de cs \\ us & \de cs & 1 + \de s^2 \end{array}\right], \label{3.3.2} \ea
where $\de = \ga - 1$.
[In this representation, $u = (\ga^2 - 1)^{1/2}$, and $c$ and $s$ can be positive or negative.]
$B(u,\th)$ has determinant 1 because it is the product of matrices with determinant 1. Notice that $B(u,\th) = B_x(u)$ and $B_y(u)$ when $\th = 0$ and $\pi/2$, respectively.

The matrix $R(\th)$ in Eq. (\ref{3.3.2}) is the inverse of the matrix in Eq. (\ref{3.2.3}), because coordinate transformations are passive rotations and $R_p(\th) = R_a^{-1}(\th) = R_a(-\th)$, where the subscripts $a$ and $p$ represent active and passive, respectively. Henceforth, we will consider only active rotations, in terms of which $B(u,\th) = R(\th)B(u)R^{-1}(\th) = R(\th)B(u)R^t(\th)$.
%{\color{green} Notice that $B(u,\th)$ is a similarity transform of $B(u)$, with the same boost magnitude, but a different boost direction.}

Formula (\ref{3.3.2}) for $B(u,\th)$ is more complicated than formulas (\ref{3.2.1}) for $B_x(u)$ and $B_y(u)$. One can determine the physical significances of some of its components by considering its effects on the vector $E_0 = [1, 0, 0]^t$, which is the energy-momentum three-vector of a particle at rest. After the boost, the three-vector
\be \left[\begin{array}{ccc} \ga & uc & us \\ uc & 1 + \de c^2 & \de cs \\ us & \de cs & 1 + \de s^2 \end{array}\right]
\left[\begin{array}{c} 1 \\ 0 \\ 0 \end{array}\right]
= \left[\begin{array}{c} \ga \\ uc \\ us \end{array}\right]. \label{3.3.3} \ee
Hence, the first column of the boost matrix is the energy-momentum three-vector that defines the transformation. The matrix is symmetric, so its first row also equals this three-vector. The spatial block is the sum of the $2 \times 2$ identity matrix and the outer product $\de [c, s]^t[c, s]$, but its physical significance remains to be determined.

It follows from the rules of matrix algebra that $B^{-1}(u,\th) = R(\th)B^{-1}(u)R^{-1}(\th) = R(\th)B(-u)R^{-1}(\th)$. One obtains the formula for $B^{-1}(u,\th)$ from the formula for $B(u,\th)$ by changing the sign of $u$ (or the signs of $c$ and $s$).
%Alternatively, one can change the signs of the components $u_x$ and $u_y$.
The inverse of a boost is a brake of the same magnitude.
%The diagonal blocks remain the same, whereas the row and column vectors change sign.
To verify this result, let $L = B^{-1}B$. Then
\be L = \left[\begin{array}{ccc} \ga & -uc & -us \\ -uc & 1 + \de c^2 & \de cs \\ -us & \de cs & 1 + \de s^2 \end{array}\right]
\left[\begin{array}{ccc} \ga & uc & us \\ uc & 1 + \de c^2 & \de cs \\ us & \de cs & 1 + \de s^2 \end{array}\right]. \label{3.3.4} \ee
Now write $L = [l_{ij}]$. Then the entries (components) of this symmetric matrix are
\ba l_{00} &= &\ga^2 - u^2(c^2 + s^2) \nonumber \\
&= &1, \\
l_{01} &= &\ga uc - uc(1 + \de c^2) - us(\de cs) \nonumber \\
&= &uc[\ga - 1 - \de(c^2 + s^2)] \nonumber \\
&= &0, \\
l_{02} &= &\ga us - uc(\de cs) - us(1 + \de s^2) \nonumber \\
&= &us[\ga - \de(c^2 + s^2) - 1] \nonumber \\
&= &0, \\
l_{11} &= &-(uc)^2 + (1 + \de c^2)^2 + (\de cs)^2 \nonumber \\
&= &-\de(\de + 2)c^2 + 1 + 2\de c^2 + \de^2c^2(c^2 + s^2) \nonumber \\
%&=&-(\ga^2 - 1)c^2 + 1 + (\ga^2 - 1)c^2 \nonumber \\
&= &1, \\
l_{12} &= &-u^2cs + (1 + \de c^2)\de cs + \de cs(1 + \de s^2) \nonumber \\
&= &cs[-\de(\de + 2) + 2\de + \de^2(c^2 + s^2)] \nonumber \\
%&= &-(\ga^2 - 1)cs + (\ga - 1)cs(\ga + 1) \nonumber \\
&= &0, \\
l_{22} &= &-(us)^2 + (\de cs)^2 + (1 + \de s^2)^2 \nonumber \\
&= &-\de(\de + 2)s^2 + 1 + 2\de s^2 + \de^2s^2(c^2 + s^2) \nonumber \\
%&= &-(\ga^2 - 1)s^2 + 1 + (\ga^2 - 1)s^2 \nonumber \\
&= &1. \ea
Hence, $L = I$, so the inverse formula is correct.

For reference, matrix (\ref{3.3.2}) can be written in the alternative form
\ba B(u_x, u_y) &= &\left[\begin{array}{ccc} \ga & u_x & u_y \\ u_x & 1 + \ep u_x^2 & \ep u_xu_y \\ u_y & \ep u_yu_x & 1 + \ep u_y^2 \end{array}\right], \label{3.3.11} \ea
where $u_x = uc$ and $u_y = us$ are momentum components, $\ga = (1 +u_x^2 + u_y^2)^{1/2}$ and $\ep = 1/(\ga + 1)$. Formula (\ref{3.3.11}) is equivalent to the vector formula (\ref{b8}). The alternative inverse matrix is obtained from matrix (\ref{3.3.11}) by changing the signs of $u_x$ and $u_y$.

One can also combine a boost matrix with two different rotation matrices. Let $B = B_x(u)$, $R_1 = R(\th_1)$, $R_2 = R(\th_2)$ and $L(u,\th_1,\th_2) = R_2BR_1^{-1}$. Then %, written explicitly,
\ba L(u,\th_1,\th_2) &= &\left[\begin{array}{ccc} 1 & 0 & 0 \\ 0 & c_2 & -s_2 \\ 0 & s_2 & c_2 \end{array}\right]
\left[\begin{array}{ccc} \ga & u & 0 \\ u  & \ga & 0 \\ 0 & 0 & 1 \end{array}\right]
\left[\begin{array}{ccc} 1 & 0 & 0 \\ 0 & c_1 & s_1 \\ 0 & -s_1 & c_1 \end{array}\right] \nonumber \\
&= &\left[\begin{array}{ccc} 1 & 0 & 0 \\ 0 & c_2 & -s_2 \\ 0 & s_2 & c_2 \end{array}\right]
\left[\begin{array}{ccc} \ga & uc_1 & us_1 \\ u  & \ga c_1 & \ga s_1 \\ 0 & -s_1 & c_1 \end{array}\right] \nonumber \\
&= &\left[\begin{array}{ccc} \ga & uc_1 & us_1 \\ uc_2  & \ga c_2c_1 + s_2s_1 & \ga c_2s_1 - s_2c_1 \\ us_2 & \ga s_2c_1 - c_2s_1 & \ga s_2s_1 + c_2c_1 \end{array}\right] \nonumber \\
&= &\left[\begin{array}{ccc} \ga & uc_1 & us_1 \\ uc_2  & c_{21} + \de c_2c_1 & -s_{21} + \de c_2s_1 \\ us_2 & s_{21} + \de s_2c_1 & c_{21} + \de s_2s_1 \end{array}\right], \label{3.3.12} \ea
where $c_{21} = \cos\th_{21}$, $s_{21} = \sin\th_{21}$ and $\th_{21} = \th_2 - \th_1$.
The angles $\th_1$, $\th_2$ and $\th_{21}$ are called the input, output and difference angles, respectively.
$L$ has determinant 1 because it is the product of matrices with determinant 1. The first column of this matrix, $[\ga, uc_2, us_2]^t$, involves the angle $\th_2$, whereas the first row, $[\ga, uc_1, us_1]$, involves the angle $\th_1$. The spatial block is the sum of the $2 \times 2$ rotation matrix $R_{21} = R(\th_{21})$ and the outer product $\de[c_2, s_2]^t [c_1, s_1]$.
One can explain the presence of the $2 \times 2$ rotation matrix by writing $R_2BR_1^{-1}$ as $(R_2R_1^{-1})(R_1BR_1^{-1})$, where $R_1BR_1^{-1}$ represents a boost in the direction $(c_1, s_1)$ and $R_2R_1^{-1} = R_{21}$ represents a rotation through the angle $\th_{21}$, and noting that the spatial block of the boost matrix contains the $2 \times 2$ identity matrix [Eq. (\ref{3.3.3})]. Alternatively, one can rewrite $R_2BR_1^{-1}$ as $(R_2BR_2^{-1})(R_2R_1^{-1})$, where $R_2BR_2^{-1}$ represents a boost in the direction $(c_2,s_2)$ and $R_2R_1^{-1}$ represents the same rotation [Eqs. (\ref{3.2.5}) and (\ref{3.2.6})].
The inverse of $R_2B(u)R_1^{-1}$ is $R_1B(-u)R_2^{-1}$. One obtains the second matrix from the first by interchanging the indices 1 and 2 (which corresponds to transposition), and changing the sign of $u$. This result is consistent with Eq. (\ref{3.1.2}).

In Sec. 3.5, we show that Eq. (\ref{3.3.12}) is the general form of a Lorentz matrix.
It is clear from the first line of Eq. (\ref{3.3.12}) that $R(\th_3)L(u, \th_1, \th_2)R^t(\th_3) = L(u, \th_1+\th_3, \th_2 + \th_3)$. The similarity transform of a Lorentz matrix is another Lorentz matrix, with the same momentum (energy), but different input and output angles. The matrices share the same difference angle.

\newpage

\subsec{3.4 Combinations of boosts}

Let $B_1 = B(u_1, \th_1)$ and $B_2 = B(u_2, \th_2)$ be matrices of the form (\ref{3.3.2}), and let $L_{21} = B_2B_1$. Then
\be L_{21} = \left[\begin{array}{ccc} \ga_2 & u_2c_2 & u_2s_2 \\ u_2c_2 & 1 + \de_2c_2^2 & \de_2c_2s_2 \\ u_2s_2 & \de_2s_2c_2 & 1 + \de_2s_2^2 \end{array}\right]
\left[\begin{array}{ccc} \ga_1 & u_1c_1 & u_1s_1 \\ u_1c_1 & 1 + \de_1c_1^2 & \de_1c_1s_1 \\ u_1s_1 & \de_1s_1c_1 & 1 + \de_1 s_1^2 \end{array}\right]. \label{3.4.1} \ee
The components of $L_{21} = [l_{ij}]$ are
\ba l_{00} &= &\ga_2\ga_1 + (u_2c_2)(u_1c_1) + (u_2s_2)(u_1s_1), \label{3.4.2} \\
l_{01} &= &\ga_2(u_1c_1) + (u_2c_2)(1 + \de_1c_1^2) + (u_2s_2)(\de_1s_1c_1), \label{3.4.3} \\
l_{02} &= &\ga_2(u_1s_1) + (u_2c_2)(\de_1c_1s_1) + (u_2s_2)(1 + \de_1s_1^2), \label{3.4.4} \\
l_{10} &= &(u_2c_2)\ga_1 + (1 + \de_2c_2^2)(u_1c_1) + (\de_2c_2s_2)(u_1s_1), \label{3.4.5} \\
l_{11} &= &(u_2c_2)(u_1c_1) + (1 + \de_2c_2^2)(1 + \de_1c_1^2) + (\de_2c_2s_2)(\de_1s_1c_1), \label{3.4.6} \\
l_{12} &= &(u_2c_2)(u_1s_1) + (1 + \de_2c_2^2)(\de_1s_1c_1) + (\de_2c_2s_2)(1 + \de_1s_1^2), \label{3.4.7} \\
l_{20} &= &(u_2s_2)\ga_1 + (\de_2s_2c_2)(u_1c_1) + (1 + \de_2s_2^2)(u_1s_1), \label{3.4.8} \\
l_{21} &= &(u_2s_2)(u_1c_1) + (\de_2s_2c_2)(1 + \de_1c_1^2) + (1 + \de_2s_2^2)(\de_1s_1c_1), \label{3.4.9} \\
l_{22} &= &(u_2s_2)(u_1s_1) + (\de_2s_2c_2)(\de_1c_1s_1) + (1 + \de_2s_2^2)(1 + \de_1s_1^2). \label{3.4.10} \ea
Whatever $L_{21}$ represents, it is not (necessarily) a boost matrix, because it is not (necessarily) symmetric. It is easy to show that the product of two symmetric matrices is symmetric if and only if the matrices commute, so a lack of symmetry means that $L_{21} \neq L_{12}$: In two space dimensions, the order of the boosts matters.

It follows from Eq. (\ref{3.3.3}) that the first column of $L_{21}$, $[l_{00}, l_{10}, l_{20}]^t$, is the energy-momentum three-vector $E_{21} = [\ga_{21}, u_{21x}, u_{21y}]^t$. In terms of components,
\ba \ga_{21} &= &\ga_2\ga_1 + u_{2x}u_{1x} + u_{2y}u_{1y}, \label{3.4.11} \\
u_{21x} &= &u_{2x}\ga_1 + (1 + \ep_2u_{2x}^2)u_{1x} + (\ep_2u_{2x}u_{2y})u_{1y} \nonumber \\
&= &u_{1x} + [\ga_1 + \ep_2(u_{2x}u_{1x} + u_{2y}u_{1y})]u_{2x}, \label{3.4.12} \\
u_{21y} &= &u_{2y}\ga_1 + (\ep_2u_{2x}u_{2y})u_{1x} + (1 + \ep_2u_{2y}^2)u_{1y} \nonumber \\
&= &u_{1y} + [\ga_1 + \ep_2(u_{2x}u_{1x} + u_{2y}u_{1y})]u_{2y}. \label{3.4.13} \ea
Equations (\ref{3.4.11}) -- (\ref{3.4.13}) are equivalent to the vector equations (\ref{b21}) and (\ref{b22}). It follows from the equation $E_{21}^tSE_{21} = E_0^tB_1^tB_2^tSB_2B_1E_0 = E_0^tB_1^tSB_1E_0 = E_0^tSE_0 = 1$ that $\ga_{21}^2 - u_{21x}^2 - u_{21y}^2 = 1$. This result is not obvious from the component equations.

It follows from the equation $(B_2B_1)^t = B_1^tB_2^t = B_1B_2$ that the first row of $L_{21}$, $[l_{00}, l_{01}, l_{02}]^t$, is the first column of $L_{12}$, which is the three-vector $E_{12} = [\ga_{12}, u_{12x}, u_{12y}]^t$. In terms of components,
\ba \ga_{12} &= &\ga_1\ga_2 + u_{1x}u_{2x} + u_{1y}u_{2y}, \label{3.4.14} \\
u_{12x} &= &u_{1x}\ga_2 + (1 + \ep_1u_{1x}^2)u_{2x} + (\ep_1u_{1x}u_{1y})u_{2y} \nonumber \\
&= &u_{2x} + [\ga_2 + \ep_1(u_{1x}u_{2x} + u_{1y}u_{2y})]u_{1x}, \label{3.4.15} \\
u_{12y} &= &u_{1y}\ga_2 + (\ep_1u_{1x}u_{1y})u_{2x} + (1 + \ep_1u_{1y}^2)u_{2y} \nonumber \\
&= &u_{2y} + [\ga_2 + \ep_1(u_{1x}u_{2x} + u_{1y}u_{2y})]u_{1y}. \label{3.4.16} \ea
Equations (\ref{3.4.14}) -- (\ref{3.4.16}) are equivalent to the vector equations (\ref{b30}) and (\ref{b31}). The components of $E_{12}$ also satisfy the identity $\ga_{12}^2 - u_{12x}^2 - u_{12y}^2 = 1$. By comparing Eqs. (\ref{3.4.11}) and (\ref{3.4.14}), one finds that $\ga_{12} = \ga_{21}$. Hence, $|\vu_{12}|^2 = |\vu_{21}|^2$. This result is also not obvious from the component equations.

The inverse of $B(u_2,\th_2)B(u_1, \th_1)$ is $B(-u_1, \th_1)B(-u_2, \th_2)$, so the inverse of two boosts is two brakes, done in the opposite order. (Undo boost 2, then undo boost 1.) Let $M_{21} = B_1^{-1}B_2^{-1}$. Then
\be M_{21} = \left[\begin{array}{ccc} \ga_1 & -u_1c_1 & -u_1s_1 \\ -u_1c_1 & 1 + \de_1c_1^2 & \de_1c_1s_1 \\ -u_1s_1 & \de_1s_1c_1 & 1 + \de_1 s_1^2 \end{array}\right]
\left[\begin{array}{ccc} \ga_2 & -u_2c_2 & -u_2s_2 \\ -u_2c_2 & 1 + \de_2c_2^2 & \de_2c_2s_2 \\ -u_2s_2 & \de_2s_2c_2 & 1 + \de_2s_2^2 \end{array}\right]. \label{3.4.17} \ee
The components of $M = [m_{ij}]$ are
\ba m_{00} &= &\ga_1\ga_2 + (u_1c_1)(u_2c_2) + (u_1s_1)(u_2s_2), \label{3.4.18} \\ 
m_{01} &= &-\ga_1(u_2c_2) - (u_1c_1)(1 + \de_2c_2^2) - (u_1s_1)(\de_2s_2c_2), \label{3.4.19} \\
m_{02} &= &-\ga_1(u_2s_2) - (u_1c_1)(\de_2c_2s_2) - (u_1s_1)(1 + \de_2s_2^2), \label{3.4.20} \\
m_{10} &= &-(u_1c_1)\ga_2 - (1 + \de_1c_1^2)(u_2c_2) - (\de_1c_1s_1)(u_2s_2), \label{3.4.21} \\
m_{11} &= &(u_1c_1)(u_2c_2) + (1 + \de_1c_1^2)(1 + \de_2c_2^2) + (\de_1c_1s_1)(\de_2s_2c_2), \label{3.4.22} \\
m_{12} &= &(u_1c_1)(u_2s_2) + (1 + \de_1c_1^2)(\de_2s_2c_2) + (\de_1c_1s_1)(1 + \de_2s_2^2), \label{3.4.23} \\
m_{20} &= &-(u_1s_1)\ga_2 - (\de_1s_1c_1)(u_2c_2) - (1 + \de_1s_1^2)(u_2s_2), \label{3.4.24} \\
m_{21} &= &(u_1s_1)(u_2c_2) + (\de_1s_1c_1)(1 + \de_2c_2^2) + (1 + \de_1s_1^2)(\de_2s_2c_2), \label{3.4.25} \\
m_{22} &= &(u_1s_1)(u_2s_2) + (\de_1s_1c_1)(\de_2s_2c_2) + (1 + \de_1s_1^2)(1 + \de_2s_2^2). \label{3.4.26} \ea
By comparing Eqs. (\ref{3.4.18}) -- (\ref{3.4.26}) with Eqs. (\ref{3.4.2}) -- (\ref{3.4.10}), one finds that
$m_{00} = l_{00}$, $m_{01} = -l_{10}$, $m_{02} = -l_{20}$, $m_{10} = -l_{01}$ and $m_{20} = -l_{02}$. Furthermore, $m_{11} = l_{11}$, $m_{12} = l_{21}$, $m_{21} = l_{12}$ and $m_{22} = l_{11}$. One obtains $M_{21} = L_{21}^{-1}$ from $L_{21}$ by transposing it and changing the signs of the momentum elements ($l_{01}$, $l_{02}$, $l_{10}$ and $l_{20}$). Notice the requirement for transposition, which is not necessary for the constituent boost matrices (because they are symmetric). This result is consistent with Eq. (\ref{3.1.2}).

When two or more transformations are considered, the algebra becomes very complicated, very quickly, and it is difficult to understand the physical significance of the mathematical results. Fortunately, there is a mathematical formalism that clarifies the structure and significance of Lorentz transformations.

\newpage

\subsec{3.5 General form of a Lorentz matrix}

What is the general form of a Lorentz matrix? Let $L = [l_{ij}]$ be a $3 \times 3$ matrix. Then, as stated in Sec. 3.1, $L$ is Lorentzian if and only if it satisfies the equivalent equations $L^tSL = S$ [Eq. (\ref{3.1.1})] and $L^{-1} = SL^tS$ [Eq. (\ref{3.1.2})], where $S$ is the metric matrix.
The Lorentz condition (\ref{3.1.1}) involves nine scalar equations for the components $l_{ij}$, but $(L^tSL)^t = L^tSL$, so only six of these equations are independent. Hence, $L$ is specified by three free parameters.
Examples of Lorentz matrices include the identity matrix (which has no free parameter), and rotation and boost matrices (which have one and two free parameters, respectively). None of these examples has three free parameters, so they are special cases of Lorentz matrices.

In this section, we derive the general form of a Lorentz matrix.
%(The same derivation is also provided in \cite{mck25c}.)}
It is convenient to write
\be L = \left[\begin{array}{cc} \ga & R \\ C & M \end{array}\right], \label{3.5.3} \ee
where $\ga$ is a scalar, $C$ is a $2 \times 1$ column vector, $R$ is a $1 \times 2$ row vector and $M$ is a $2 \times 2$ matrix. It follows from Eqs. (\ref{3.1.2}) and (\ref{3.5.3}) that
\be L^{-1} = \left[\begin{array}{cc} \ga & -C^t \\ -R^t & M^t \end{array}\right]. \label{3.5.4} \ee
Formula (\ref{3.5.4}) is consistent with the examples of Secs. 3.3 and 3.4 (transpose the matrix and change the signs of the column and row vectors).

By combining Eqs. (\ref{3.5.3}) and (\ref{3.5.4}), one finds that
\ba \left[\begin{array}{cc} \ga & -C^t \\ -R^t & M^t \end{array}\right]
\left[\begin{array}{cc} \ga & R \\ C & M \end{array}\right]
&= &\left[\begin{array}{cc} \ga^2 - C^tC & \ga R - C^tM \\ M^tC - \ga R^t & M^tM - R^tR \end{array}\right], \label{3.5.5} \\
\left[\begin{array}{cc} \ga & R \\ C & M \end{array}\right]
\left[\begin{array}{cc} \ga & -C^t \\ -R^t & M^t \end{array}\right]
&= &\left[\begin{array}{cc} \ga^2 - RR^t & RM^t - \ga C^t \\ \ga C - MR^t & MM^t - CC^t \end{array}\right]. \label{3.5.6} \ea
The matrices on the right sides of Eqs. (\ref{3.5.5}) and (\ref{3.5.6}) should equal the identity matrix $I$.

It follows from the top-left entries of Eqs. (\ref{3.5.5}) and (\ref{3.5.6}) that
\be \ga^2 - 1 = C^tC = RR^t. \label{3.5.7} \ee
Hence, $C$ and $R$ have the same length, $u = (\ga^2 - 1)^{1/2}$, in which case $C = u[\cos(\th_2), \sin(\th_2)]^t$ and $R =u [\cos(\th_1), \sin(\th_1)]$.

The scalar $\ga$, and the vectors $C$ and $R$, are specified by three parameters, namely $\ga$ itself, and the angles $\th_1$ and $\th_2$. No free parameters remain, so $M$ must be specified by scalar functions of $\ga$ and matrix combinations of $C$ and $R$. The top-right entry of Eq. (\ref{3.5.5}) requires that $C^tM = \ga R$ and the bottom-left entry of Eq. (\ref{3.5.6}) requires that $MR^t = \ga C$. $M$ is a rotation-like matrix, because it converts $R^t$ to $C$ and $C^t$ to $R$ [Eqs. (\ref{3.2.5}) and (\ref{3.2.6})], but it is not a rotation matrix because $\ga \neq 1$. Of the outer products $CC^t$, $R^tR$, $CR$ and $R^tC^t$, only the third has the property that $C^t(CR) \propto R$ and $(CR)R^t \propto C$. Hence, we choose the ansatz $M = N + \ep CR$, where the matrix $N$ and scalar $\ep$ remain to be determined.

The bottom-left entry of Eq. (\ref{3.5.6}) is
\be 0 = \ga C - MR^t = \ga C - (NR^t + \ep u^2 C). \label{3.5.8} \ee
This equation, which involves arbitrary vectors, requires that $NR^t = C$, so $N$ is the rotation matrix that converts $R^t$ to $C$. The subsequent equation $\ga - 1 = \ep u^2$ requires that $\ep = 1/(\ga + 1)$. With $N$ and $\ep$ so defined, the ansatz satisfies the bottom-left equation.

The top-right entry of Eq. (\ref{3.5.5}) is
\be 0 = \ga R - C^tM = \ga R - (C^tN + \ep u^2 R). \label{3.5.9} \ee
The identity $NR^t = C$ implies that $C^tN = RN^tN = R$, as required, and the identity $\ep u^2$ $= \ga - 1$ ensures that the top-right equation is satisfied.

The bottom-right entry of matrix (\ref{3.5.6}) is
\ba MM^t - CC^t &= &(N + \ep CR)(N^t + \ep R^tC^t) - CC^t \nonumber \\
&= &I + \ep CC^t + \ep CC^t + \ep^2u^2 CC^t - CC^t. \label{3.5.10} \ea
The $CC^t$ terms on the right side cancel, because $\ep(2 + \ep u^2) = 1$, so $MM^t - CC^t = I$, as required. The $R^tR$ terms in the bottom-right entry of matrix (\ref{3.5.5}) cancel for the same reason.

The preceding results are summarized by the equations
\ba \left[\begin{array}{cc} \ga & R \\ C & N + \ep CR \end{array}\right]
&= &\left[\begin{array}{cc} 1 & 0 \\ 0 & N \end{array}\right]
\left[\begin{array}{cc} \ga & R \\ R^t & I + \ep R^tR \end{array}\right] \nonumber \\
&= &\left[\begin{array}{cc} \ga & C^t \\ C & I + \ep CC^t \end{array}\right]
\left[\begin{array}{cc} 1 & 0 \\ 0 & N \end{array}\right]. \label{3.5.11} \ea
Thus, a Lorentz transformation consists of a boost followed by a rotation, or the same rotation followed by what appears to be a different boost. The free parameters are the energy (or momentum) and direction angle of the boost, and the rotation angle (QED).

Notice that the first matrix in Eqs. (\ref{3.5.11}) has the same form as matrix (\ref{3.3.12}). This result is important. Let $L = [l_{ij}]$. Then it follows from Eqs. (\ref{3.3.12}) and (\ref{3.5.11}) that the energy parameter
\be \ga = l_{00}, \label{3.5.12} \ee
and the input and output angles are specified implicitly by the equations
\be \tan(\th_1) = l_{02}/l_{01}, \ \ \tan(\th_2) = l_{20}/l_{10}. \label{3.5.13} \ee
Furthermore, $l_{11} + l_{22} = (\ga + 1)\cos(\th_{21})$ and $l_{21} - l_{12} = (\ga + 1)\sin(\th_{21})$, where $\th_{21} = \th_2 - \th_1$, so the difference angle is specified by the equation
\be \tan(\th_{21}) = (l_{21} - l_{12})/(l_{11} + l_{22}). \label{3.5.14} \ee
When one is presented with a Lorentz matrix, one can use Eqs. (\ref{3.5.12}) -- (\ref{3.5.14}) to determine the energy parameter, and the input, output and difference angles, respectively.

Although we chose to focus on active transformations, it is sometimes helpful to think of passive transformations. Consider two pairs of coordinate axes, one in the laboratory frame (LF) and the other in a moving frame (MF), which moves with velocity $\vv$ relative to the LF. At some reference time, the two origins coincide. How does one transform the LF axes so that they coincide with the MF axes? First, one boosts the LF axes so that their origin coincides (keeps up) with the origin of the MF axes. Second, one rotates the LF axes so that they align with the MF axes. Thus, the transformation between the frames is a boost followed by a rotation \cite{wig39}. Alternatively, one could rotate the LF axes first, then boost them until their origin coincides with the origin of the MF axes. In this passive picture, it is clear that the two transformations involve the same boost. The boost matrices in Eq. (\ref{3.5.11}) are different because the boost is described in different coordinate systems (relative to unrotated and rotated axes). This physical interpretation is consistent with the mathematics: If $L = RB_1 = B_2R$, then $B_2 = RB_1R^t$. This relation is a similarity transformation, which preserves the boost energy.

In the preceding analysis, $R$ and $C$ ($\th_1$ and $\th_2$) were chosen freely, and $N$ ($\th_{21}$) and $M = N + \ep CR$ were specified. It is also possible to choose $R$ and $N$ ($\th_1$ and $\th_{21}$) freely, in which case $C = NR^t$ ($\th_2$) and $M = N + \ep CR = N(I + \ep R^tR)$ are specified.

In three dimensions, the Lorentz condition (\ref{3.1.1}) represents 16 scalar equations, but because of symmetry, only 10 are independent. Hence, $L$ is specified by six free parameters. Formula (\ref{3.5.3}) remains valid (although the dimensions of $C$, $R$ and $M$ change). So also does the proof that the transformation consists of a boost followed by a rotation, or the same rotation followed by a boost [Eqs. (\ref{3.5.11})]. Hence, the free parameters are the momentum components of the boost (or the energy of the boost and two polar angles that specify its direction) and the angles that specify the rotation (two polar angles that specify its axis and one rotation angle).

\newpage

\subsec{3.6 Schmidt decomposition}

Let $M$ be a real matrix. Then $M$ has the Schmidt decomposition $QDP^t$, where $D$ is a non-negative diagonal matrix, and $P$ and $Q$ are orthogonal matrices \cite{hor13}. The columns of $P$ (input Schmidt vectors) are the eigenvectors of $M^tM$, the columns of $Q$ (output Schmidt vectors) are the eigenvectors of $MM^t$, and the components of $D$ (Schmidt coefficients) are the square roots of the (common) eigenvalues of $M^tM$ and $MM^t$. If $E$ is an input vector, then $F = ME$ is the associated output vector, and conversely, if $F$ is an output vector, then $E = M^tF$ is the associated input vector. Schmidt decompositions arise in a variety of fields, including nonlinear and quantum optics \cite{eke95,law00,mck13a,mck13b}.

Now let $L$ be a Lorentz matrix. Then $L^tL$ is also a (symmetric) Lorentz matrix. Suppose that $(L^tL)E = \la E$, in which case $E$ is an eigenvector of $L^tL$ (and an input vector of $L$), with the associated eigenvalue (squared coefficient) $\la$. Then $E^tL^tLE = (LE)^t(LE) = \la E^tE$. In this equation, the inner products are non-negative, so the eigenvalue is also non-negative. The product of the eigenvalues of $L^tL$ equals its determinant [Eq. (\ref{3.8.11})], which is 1. Hence, $L^tL$ cannot have a zero eigenvalue, so its eigenvalues are positive. If $E$ is an eigenvalue of $L^tL$ with eigenvalue $\la$, then $E$ is also an eigenvector of $(L^tL)^{-1} = S(L^tL)S$ with eigenvalue $\la^{-1}$. By multiplying the eigenvalue equation by $S$ on the left, one finds that $L^tL(SE) = \la^{-1}(SE)$. If $E$ is an eigenvector of $L^tL$ with eigenvalue $\la$, then $SE$ is also an eigenvector with eigenvalue $\la^{-1}$. The eigenvalues of $L^tL$ occur in reciprocal pairs. A $3 \times 3$ matrix has one pair of reciprocal eigenvalues, and one unit eigenvalue, which is its own reciprocal.
Similar statements can be made about $LL^t$ and its eigenvector $F$ (which is an output vector of $L$).
To summarize these results, every Lorentz matrix has a pair of reciprocal Schmidt coefficients and a unit coefficient.

The block form of matrix (\ref{3.5.11}) facilitates studies of its Schmidt decomposition. In the symmetric case ($C = R^t$), Eq. (\ref{3.5.11}) reduces to
\be L = \left[\begin{array}{cc} \ga & C^t \\ C & I + \ep CC^t \end{array}\right]. \label{3.6.1} \ee
Let $B$ be a $2 \times 1$ vector that is orthogonal to $C$ ($B^tC = 0 = C^tB$). Then it is easy to verify that the eigenvectors of $L$ are
\be {1 \over 2^{1/2}}\left[\begin{array}{c} 1 \\ C/u \end{array}\right], \ \ 
{1 \over 2^{1/2}}\left[\begin{array}{c} 1 \\ -C/u \end{array}\right], \ \ 
\left[\begin{array}{c} 0 \\ B \end{array}\right], \label{3.6.2} \ee
and the corresponding eigenvalues are
\be \ga + u, \ \ \ga - u, \ \ 1, \label{3.6.3} \ee
respectively. The auxiliary equation ensures that $\ga - u = 1/(\ga + u)$. The spectral decomposition of a symmetric matrix with positive eigenvalues is identical to its Schmidt decomposition.

In the asymmetric case,
\be L = \left[\begin{array}{cc} \ga & R \\ C & N + \ep CR \end{array}\right]. \label{3.6.4} \ee
The first product matrix
\be L^tL = \left[\begin{array}{cc} \ga^2 + u^2 & 2\ga R \\ 2\ga R^t & I + 2R^tR \end{array}\right]. \label{3.6.5} \ee
It is easy to verify that the input Schmidt vectors ($E_i$) are
\be {1 \over 2^{1/2}}\left[\begin{array}{c} 1 \\ R^t/u \end{array}\right], \ \ 
{1 \over 2^{1/2}}\left[\begin{array}{c} 1 \\ -R^t/u \end{array}\right], \ \ 
\left[\begin{array}{c} 0 \\ B \end{array}\right], \label{3.6.6} \ee
where $RB = 0$, and the associated (squared) Schmidt coefficients ($\si_i^2$) are
\be (\ga + u)^2, \ \ (\ga - u)^2, \ \ 1, \label{3.6.7} \ee
respectively. The second product matrix
\be LL^t =  \left[\begin{array}{cc} \ga^2 + u^2 & 2\ga C^t \\ 2\ga C & I + 2CC^t \end{array}\right]. \label{3.6.8} \ee
It is easy to verify that the output vectors ($F_i$) are
\be {1 \over 2^{1/2}}\left[\begin{array}{c} 1 \\ C/u \end{array}\right], \ \ 
{1 \over 2^{1/2}}\left[\begin{array}{c} 1 \\ -C/u \end{array}\right], \ \ 
\left[\begin{array}{c} 0 \\ NB \end{array}\right], \label{3.6.9} \ee
where $C^tNB = RN^tNB = RB = 0$, and the associated (squared) coefficients are
\be (\ga + u)^2, \ \ (\ga - u)^2, \ \ 1, \label{3.6.10} \ee
respectively.

In terms of input and output vectors, $L = \tsum_i F_i\si_iE_i^t$. Hence, one can multiply any pair of vectors by the same phase factor without changing the decomposition. We chose to multiply the second pair by $-1$.
Written explicitly,
%2
\be L = \left[\begin{array}{ccc} r & -r & 0 \\ rc_2 & rc_2 & -s_2 \\ rs_2 & rs_2 & c_2 \end{array}\right]
\left[\begin{array}{ccc} \ga + u & 0 & 0 \\ 0 & \ga - u & 0 \\ 0 & 0 & 1 \end{array}\right]
\left[\begin{array}{ccc} r & rc_1 & rs_1 \\ -r & rc_1 & rs_1 \\ 0 & -s_1 & c_1 \end{array}\right], \label{3.6.11} \ee
where $r = 1/2^{1/2}$. Multiplying the matrices on the right side of Eq. (\ref{3.6.11}) produces matrix (\ref{3.3.12}). It is easy to verify that
\be \left[\begin{array}{cc} r & -r \\ r & r \end{array}\right]
\left[\begin{array}{cc} \ga + u & 0 \\ 0 & \ga - u \end{array}\right]
\left[\begin{array}{cc} r & r \\ -r & r \end{array}\right]
= \left[\begin{array}{cc} \ga & u \\ u & \ga \end{array}\right]. \label{3.6.12} \ee
Let $R_{tx}$ and $B_x$ be the first and last matrices in Eq. (\ref{3.6.12}), respectively. Then Eq. (\ref{3.6.11}) can be rewritten as $L = (QR_{tx}^t)(R_{tx}DR_{tx}^t)(R_{tx}P^t)$, where the middle matrix represents a boost in the $x$-direction and the other matrices are orthogonal. (In this equation, $R_{tx}$ is the $3 \times 3$ extension of the $2 \times 2$ matrix defined above.) Written explicitly,
\ba QR_{tx}^t &= &\left[\begin{array}{ccc} r & -r & 0 \\ rc_2 & rc_2 & -s_2 \\ rs_2 & rs_2 & c_2 \end{array}\right]
\left[\begin{array}{ccc} r & r & 0 \\ -r & r & 0 \\ 0 & 0 & 1 \end{array}\right]
\ = \ \left[\begin{array}{ccc} 1 & 0 & 0 \\ 0 & c_2 & -s_2 \\ 0 & s_2 & c_2 \end{array}\right], \label{3.6.13} \\
R_{tx}P^t &= &\left[\begin{array}{ccc} r & -r & 0 \\ r & r & 0 \\ 0 & 0 & 1 \end{array}\right]
\left[\begin{array}{ccc} r & rc_1 & rs_1 \\ -r & rc_1 & rs_1 \\ 0 & -s_1 & c_1 \end{array}\right]
\ = \ \left[\begin{array}{ccc} 1 & 0 & 0 \\ 0 & c_1 & s_1 \\ 0 & -s_1 & c_1 \end{array}\right]. \label{3.6.14} \ea
Matrices (\ref{3.6.13}) and (\ref{3.6.14}) represent two-dimensional rotations (rather then three-dimensional orthogonal transformations). The Schmidt-like decomposition $L = Q_2B_xP_2^t$ is even more informative and useful than the Schmidt decomposition, because $P_2 = PR_{tx}^t$ and $Q_2 = QR_{tx}^t$ represent simple rotations.

If one regards the first matrix on the left side of Eq. (\ref{3.6.12}) as an active matrix, which rotates basis vectors by $\pi/4$ radians, then the third matrix is the associated passive matrix, which rotates coordinates. Hence,
\be  \left[\begin{array}{c} t' \\ x' \end{array}\right]
= \left[\begin{array}{cc} r & r \\ -r & r \end{array}\right]
\left[\begin{array}{c} t \\ x \end{array}\right]
= \left[\begin{array}{c} r(t + x) \\ r(x - t) \end{array}\right]. \label{3.6.15} \ee
The last vector in Eq. (\ref{3.6.15}) involves the characteristic variables $\tau = r(t + x)$ and $\xi = r(x - t)$. (The other characteristic variable $\et = y$.) If light is emitted at the origin, its leading edge is specified by $\xi = 0$, whereas its trailing edge is specified by $\ta = 0$. According to Eq. (\ref{3.6.12}), $\ta$ is stretched by the factor $\ga + u$, whereas $\xi$ is squeezed by the same factor (and $\et$ is unaffected). These variables are well defined. However, we prefer the alternative characteristic variables $\ta = r(t - x)$ and $\xi = r(x + t)$, the first of which is called the retarded time \cite{kim82}. If light is emitted at the origin, its leading edge is specified by $\ta = 0$, whereas its trailing edge is specified by $\xi = 0$. Both pairs of characteristic variables are physically meaningful, but only the first pair appeared naturally.

In the Schmidt decomposition $L = \tsum_i F_i\si_iE_i^t$, one can do the summation in any order without affecting the decomposition. Interchanging the first and second vectors in Eqs. (\ref{3.6.6}) and (\ref{3.6.9}) correponds to interchanging the first and second coefficients ($\ga + u$ and $\ga - u$). In Eq. (\ref{3.6.11}), the first two columns of the output matrix become $[r, -rc_2, -rs_2]^t$ and $[r, rc_2, rs_2]^t$, and the first two rows of the input matrix become $[r, -rc_1, -rs_1]$ and $[r, rc_1, rs_1]$. The matrix product does not change. In Eq. (\ref{3.6.12}) the boost matrix has the alternative decomposition
\be \left[\begin{array}{cc} r & r \\ -r & r \end{array}\right]
\left[\begin{array}{cc} \ga - u & 0 \\ 0 & \ga + u \end{array}\right]
\left[\begin{array}{cc} r & -r \\ r & r \end{array}\right]
= \left[\begin{array}{cc} \ga & u \\ u & \ga \end{array}\right]. \label{3.6.16} \ee
If one regards the first matrix on the left side of Eq. (\ref{3.6.16}) as an active matrix, then the third matrix is the associated passive matrix. Hence,
\be  \left[\begin{array}{c} t' \\ x' \end{array}\right]
= \left[\begin{array}{cc} r & -r \\ r & r \end{array}\right]
\left[\begin{array}{c} t \\ x \end{array}\right]
= \left[\begin{array}{c} r(t - x) \\ r(x + t) \end{array}\right]. \label{3.6.17} \ee
In the alternative approach, the second pair of characteristic variables arises naturally. It remains true that $P_2$ and $Q_2$ represent rotations.

In three dimensions, the Schmidt decomposition of $L$ is similar to the preceding one. The main difference is that there are four Schmidt coefficients, $\ga + u$, $\ga - u$, 1 and 1, because there are two directions, specified by the input vectors $B_1$ and $B_2$, which are perpendicular to the boost axis [Eqs. (\ref{3.6.6})]. The associated output vectors are $NB_1$ and $NB_2$ [Eqs. (\ref{3.6.9})]. $L$ also has the Schmidt-like decomposition $Q_3B_xR_3^t$, in which $B_x$ represents a boost, and $P_3$ and $Q_3$ represent three-dimensional rotations (rather than four-dimensional orthogonal transformations).

\newpage

\subsec{3.7 Spectral decomposition}

In this section, we discuss the spectral decomposition of a Lorentz matrix. Let $M$ be an invertible matrix and suppose that $ME = \la E$, where $E$ is a vector and $\la$ is a scalar. Then $M^{-1}E = \la^{-1}E$. If $E$ is an eigenvector of $M$ with the associated eigenvalue $\la$, then $E$ is also an eigenvector of $M^{-1}$ with eigenvalue $\la^{-1}$. Now let $L$ be a Lorentz matrix. Then $L^{-1}E = SL^tSE = \la^{-1}E$, where $S$ is the metric matrix. By multiplying this eigenvalue equation by $S$ on the left, one finds that $L^t(SE) = \la^{-1}(SE)$. If $E$ is an eigenvector of $L$ with eigenvalue $\la$, then $SE$ is an eigenvalue of $L^t$ with eigenvalue $\la^{-1}$. But $L$ and $L^t$ have the same eigenvalues, so $\la^{-1}$ is also an eigenvalue of $L$. Thus, the eigenvalues of $L$ occur in reciprocal pairs. For a $3 \times 3$ matrix, the third eigenvalue must be 1, which is its own reciprocal. These results are not obvious [Eq. (\ref{3.3.12})], despite their similarity to the corresponding results for Schmidt coefficients (Sec. 3.6). If $L$ is symmetric, then the eigenvectors of $L$ are orthogonal. Otherwise, they are not necessarily orthogonal.

It is well know that the sum of the eigenvalues of a matrix equals its trace, and the product of its eigenvalues equals its determinant. For matrix (\ref{3.3.12}), $\la_0 + \la_1 + \la_2 = \ga + (\ga + 1)c_{21}$ and $\la_0\la_1\la_2 = 1$, where $\ga$ is the energy and $c_{21} = \cos(\th_{21})$. The difference angle $\th_{21} = \th_2 - \th_1$, where $\th_1$ and $\th_2$ are the input and output angles, respectively. Knowing that $\la_0 = 1$ allows one to write $\la_1 + \la_2 = 2\ta$ and $\la_2\la_2 = 1$, where $\ta = [\ga - 1 + (\ga + 1)c_{21}]/2$. By solving these equations, one finds that
\be \la_{1,2} = \ta \pm (\ta^2 - 1)^{1/2} = \la_\pm. \label{3.7.1} \ee
Notice that $\la_0$ does not depend on $\th_1$ or $\th_2$, and $\la_\pm$ only depends on $\th_{21}$. For the special case in which $c_{21} = 1$ (no rotation), $\ta = \ga$ and $\la_\pm = \ga \pm u$. For the complementary case in which $\ga = 1$ (no boost), $\ta = c_{21}$ and $\la_\pm = c_{21} \pm is_{21}$, where $s_{21} = \sin(\th_{21})$.
Thus, the eigenvalues of a Lorentz matrix can be real and non-negative, or complex, with modulus~1.

The eigenvalues of a Lorentz matrix are plotted as functions of the difference angle in Fig. 1. For smaller angles ($\th_2 \ap \th_1$), the eigenvalues are real (and the transformation is boost-like), whereas for larger angles ($|\th_2 - \th_1| \sim 1$), they are complex (and the transformation is rotation-like). The transition between the regimes occurs when $\ta = 1$, or, equivalently, when $c_{21} = (3 - \ga)/(\ga + 1)$.

\newpage

\begin{figure}[t!]
\vspace*{0.0in} %-0.1
\centerline{\includegraphics[width=3in]{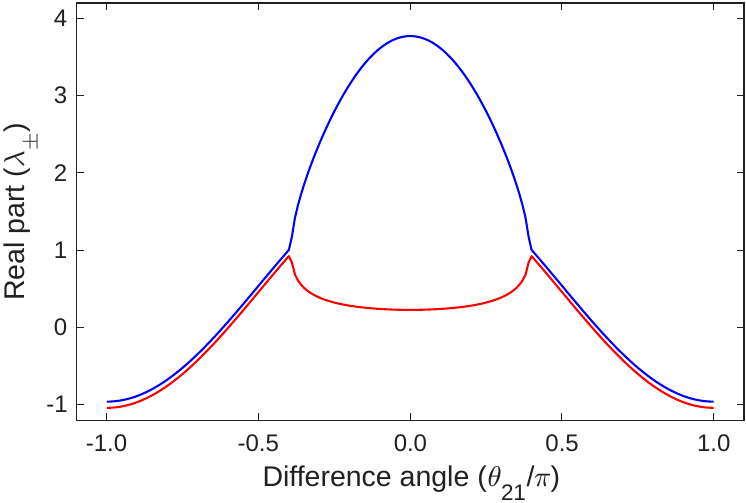} \hspace*{0.2in} \includegraphics[width=3in]{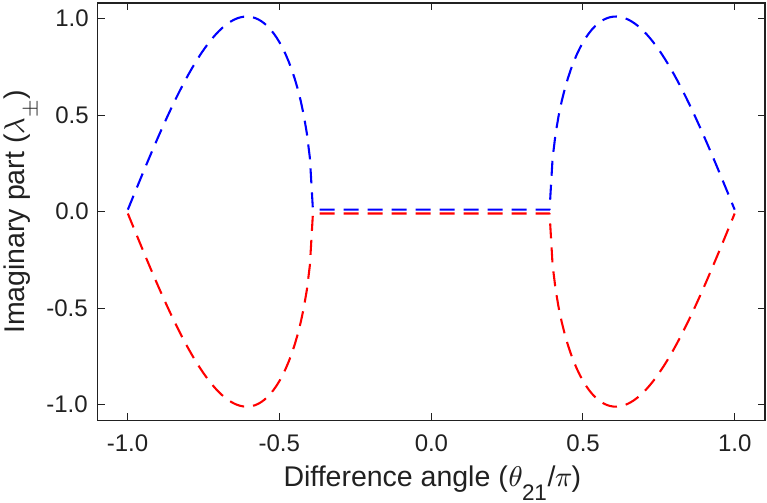}}
\vspace*{-0.1in} %-0.2
\caption{Eigenvalues of $L$ plotted as a function of $\th_{21}/\pi$ for $\ga = 2$ [Eq. (\ref{3.7.1})]. The plus (blue) and minus (red) curves were displaced slightly for visibility.}
\end{figure}

The preceding analysis provided limited physical insight into the eigenvalues and did not address the associated eigenvectors, so an alternative approach is required.
In Sec. 3.6, we showed that every Lorentz matrix has the Schmidt-like decomposition $L = QBP^t$, where $B$ represents a boost in the $x$ direction, and $P$ and $Q$ represent rotations in space through the angles $\th_1$ and $\th_2$, respectively. Consider the boost matrix
\be B = \left[\begin{array}{ccc} \ga & u & 0 \\ u & \ga & 0 \\ 0 & 0 & 1 \end{array}\right], \label{3.7.2} \ee
where $u$ is the momentum and $\ga = (1 + u^2)^{1/2}$ is the energy. It is easy to verify that $B$ has the eigenvalues
\be \la_0 = 1, \ \ \la_\pm = \ga \pm u, \label{3.7.3} \ee
and the associated eigenvectors
\be E_0 = \left[\begin{array}{c} 0 \\ 0 \\ 1 \end{array}\right], \ \ 
E_\pm = \left[\begin{array}{c} \pm r \\ r \\ 0 \end{array}\right], \label{3.7.4} \ee
where $r = 1/2^{1/2}$. Notice that the $y$ coordinate (or momentum component) is not affected by a boost in the $x$ direction. Notice also that the eigenvalues $\la_\pm$ are real, and the eigenvectors $E_\pm$ are real and orthogonal (because $B$ is real and symmetric). The eigenvectors are inclined at $\pi/4$ radians to the $t$ and $x$ axes \cite{kim82}. If $u > 0$, the characteristic coordinate associated with $E_+$ is stretched, whereas the coordinate associated with $E_-$ is squeezed.

Let $E_i$ and $E_j$ be eigenvectors of $B$ ($E_0$ or $E_\pm$). Then it is easy to verify that $E_i^tE_j = \de_{ij}$, where $\de_{ij}$ is the Kronecker delta. It is also easy to verify that $\tsum_i E_iE_i^t = I$: The set of eigenvectors is orthonormal and complete. Hence, the boost matrix has the spectral decomposition $\tsum_i E_i\la_iE_i^t = O\La O^t$, where $O$ is the orthogonal matrix whose $i$th column is $E_i$ and $\La$ is the diagonal matrix whose $i$th diagonal entry is $\la_i$. Written explicitly,
\be B = \left[\begin{array}{ccc} r & -r & 0 \\ r & r & 0 \\ 0 & 0 & 1 \end{array}\right]
\left[\begin{array}{ccc} \ga + u & 0 & 0 \\ 0 & \ga - u & 0 \\ 0 & 0 & 1 \end{array}\right]
\left[\begin{array}{ccc} r & r & 0 \\ -r & r & 0 \\ 0 & 0 & 1 \end{array}\right]. \label{3.7.5} \ee

Now consider the rotation matrix
\be R = \left[\begin{array}{ccc} 1 & 0 & 0 \\ 0 & c & -s \\ 0 & s & c \end{array}\right], \label{3.7.6} \ee
where $c = \cos(\th)$, $s = \sin(\th)$ and $\th$ is the rotation angle. It is easy to verify that $R$ has the eigenvalues
\be \la_0 = 1, \ \ \la_\pm = c \pm is = e^{\pm i\th}, \label{3.7.7} \ee % \exp(\pm i\th)
and the associated eigenvectors
\be E_0 = \left[\begin{array}{c} 1 \\ 0 \\ 0 \end{array}\right], \ \ 
E_\pm = \left[\begin{array}{c} 0 \\ r \\ \mp ir \end{array}\right]. \label{3.7.8} \ee
Notice that the time coordinate (or energy) is not affected by a spatial rotation. Notice also that the eigenvalues $\la_\pm$ and eigenvectors $E_\pm$ are complex (because $R$ is asymmetric). The real and imaginary parts of $E_\pm$ are orthogonal.

Now let $E_i$ and $E_j$ be eigenvectors of $R$ ($E_0$ or $E_\pm$). Then it is easy to verify that $E_i^\d E_j = \de_{ij}$ and $\tsum_i E_iE_i^\d = I$: Once again, the set of eigenvectors is orthonormal and complete. Hence, the rotation matrix has the spectral decomposition $\tsum_i E_i\la_iE_i^\d = U\La U^\d$, where $U$ is unitary and $\La$ is complex diagonal. Written explicitly,
\be \left[\begin{array}{ccc} 1 & 0 & 0 \\ 0 & r & r \\ 0 & -ir & ir \end{array}\right]
\left[\begin{array}{ccc} 1 & 0 & 0 \\ 0 & c + is & 0 \\ 0 & 0 & c - is \end{array}\right]
\left[\begin{array}{ccc} 1 & 0 & 0 \\ 0 & r & ir \\ 0 & r & -ir \end{array}\right]. \label{3.7.9} \ee

The preceding results can be generalized. Let $L = PBP^t$ be an arbitrary boost matrix. Then, by multiplying the fundamental eigenvalue equation $BE = \la E$ by $P$ and inserting $I = P^tP$, one finds that $PBP^tPE = L(PE) = \la(PE)$. Thus, every boost matrix has the same eigenvalues as the fundamental matrix (1, $\ga + u$ and $\ga - u$) and its eigenvectors are rotated versions of the fundamental eigenvectors ($PE$). Notice that the eigenvectors $PE_\pm$ are inclined at $\pi/4$ radians to the parallel and perpendicular boost axes.

Now let $L = QBP^t$ be an arbitrary transformation matrix, and let $\la$ and $E$ be an eigenvalue and eigenvector of $L$ (not $B$), respectively. Then $LE = QBP^tE = \la E = \la PP^tE$. By multiplying the eigenvalue equation by $Q^t$, one obtains the transformed equation $BE' = \la R^t E'$, where $E' = P^tE$ and $R = P^tQ$. This approach involves the fundamental boost matrix and the nontrivial weight matrix $R^t$. (In eigenvalue equations for differential operators, weight functions are common.) Alternatively, by multiplying the eigenvalue equation by $P^t$, one obtains the transformed equation $RBE'= \la IE'$. This approach involves a rotated boost matrix and the trivial weight matrix $I$. Notice that the transformed equations are equivalent. (Multiplying the first equation by $R$ produces the second.) In two dimensions, rotation matrices commute, so $R = QP^t$, which only depends on the difference angle.

We chose to implement the second approach, because it allows the sum and product of the eigenvalues to be $\tr(RB)$ and 1, respectively.
The characteristic matrix
\be \la I - RB = \left[\begin{array}{ccc} \la - \ga & -u & 0 \\ -uc & \la - \ga c & s \\ -us & -\ga s & \la - c \end{array}\right], \label{3.7.21} \ee
where $c = \cos(\th_{21})$ and $s = \sin(\th_{21})$, from which it follows that
\ba |\la I - RB| &= &(\la - \ga)[(\la - \ga c)(\la - c) + \ga s^2]+ u[us^2 - uc(\la - c)] \nonumber \\
&= &(\la - \ga)[\la^2 - (\ga + 1)c\la + \ga] - u^2c\la + u^2 \nonumber \\
&= &\la^3 - [\ga + (\ga + 1)c]\la^2 + [\ga + \ga(\ga + 1)c - u^2c]\la - \ga^2 + u^2 \nonumber \\
&= &\la^3 - [\ga + (\ga + 1)c]\la^2 + [\ga + (\ga + 1)c]\la - 1. \label{3.7.22} \ea
The sum of the eigenvalues is $\ga + (\ga + 1)c$ and the product is 1, as stated above. It is easy to verify that $\la_0 = 1$ is an eigenvalue. [The first and fourth terms in Eq. (\ref{3.7.22}) cancel, as do the second and third terms.] 

Let $E' = [t, x, y]^t$ be an eigenvector. Then
\be \left[\begin{array}{ccc} \la - \ga & -u & 0 \\ -uc & \la - \ga c & s \\ -us & -\ga s & \la - c \end{array}\right]
\left[\begin{array}{c} t \\ x \\ y \end{array}\right] = 0,  \label{3.7.23} \ee
from which it follows that
\be x = {\la - \ga \over u}t, \ \ 
y = {ust + \ga sx \over \la - c} = {s(\la\ga - 1) \over u(\la - c)}t. \label{3.7.24} \ee
Alternatively,
\be x = {(\la - \ga)(\la - c) \over s(\la\ga - 1)}y, \ \ 
t = {u \over \la - \ga}x = {u(\la - c) \over s(\la\ga - 1)}y. \label{3.7.25} \ee
The alternative formulas are more useful when $\th_{21} \ll 1$ and $t$, $x \ll y$ in the eigenvector $E_0'$. With $E'$ known, $E = PE'$.

Consider the direct eigenvalue problem $ME = \la E$. If $M$ is hermitian ($M^\d = M$), then the eigenvalues are real and the eigenvectors are orthogonal. This result applies to boost matrices, which are real and symmetric (so the eigenvectors are also real). However, transformation matrices can be asymmetric, in which case the eigenvalues and eigenvectors are not necessarily real, and the eigenvectors are not necessarily orthogonal. One must also consider the adjoint eigenvalue problem $M^\d F = \la^* F$ \cite{mck25a}. The eigenvalues of $M$ and $M^\d$ are conjugates of each other, but the eigenvectors are different. It follows from the problem statements that $F_2^\d ME_1 = \la_1F_2^\d E_1$ and $E_1^\d M^\d F_2 = \la_2^*E_1^\d F_2$. By combining the first equation with the hermitian conjugate of the second, one finds that $(\la_2 - \la_1)F_2^\d E_1 = F_2^\d ME_1 - F_2^\d ME_1 = 0$. Thus, the direct and adjoint eigenvectors associated with different eigenvalues are orthogonal, and one can choose their normalizations and phase factors in such a way that $F_i^\d E_i = 1$. If the eigenvectors are complete, then $M = \tsum_i E_i\la_iF_i^\d = U\La V^\d$, where $U$ ($V$) is the matrix whose $i$th column is $E_i$ ($F_i$) and $\La$ is the diagonal matrix whose $i$th diagonal entry is $\la_i$. Notice that $E_i$ appears on the left of $F_i^\d$ in the summation.

If the real matrix M has a real eigenvalue $\la$, then $ME = \la E$. The eigenvector, or its associated characteristic coordinate, is dilated (stretched or squeezed). In contrast, if $M$ has a pair of conjugate eigenvalues, then $ME = \la E$ and $ME^*= \la^* E$. In terms of real and imaginary parts, $ME_r = \la_rE_r - \la_iE_i$ and $ME_i = \la_iE_r + \la_rE_i$. The real and imaginary parts of the eigenvector are dilated and rotated. $E_r$ and $E_i$ are not necessarily orthogonal, so they are not necessarily dilated or rotated by the same amounts (but for a pure rotation, they are).

Lorentz matrices are real, so in the preceding analysis, $L^\d = L^t$. It was shown above that $L$ has three eigenvalues, $\la_0  = 1$, $\la_1$ and $\la_2 = 1/\la_1$. If $\la_1$ is real, then $\la_1 > 1$ and $\la_2 < 1$, whereas if $\la_1$ is complex, then $\la_2 = \la_1^*$ and $|\la_1|^2 = |\la_2|^2 = 1$. It was also shown that if $E_0$, $E_1$ and $E_2$ are eigenvectors of $L$, with eigenvalues $\la_0$, $\la_1$ and $\la_2$, respectively, then $SE_0$, $SE_1$ and $SE_2$ are eigenvectors of $L^t$ with eigenvalues $\la_0$, $1/\la_1 = \la_2$ and $1/\la_2 = \la_1$, respectively. It is always true that the adjoint eigenvector $F_0 = SE_0$. If $\la_1$ and $\la_2$ are real, then $F_1 = SE_2$ and $F_2 = SE_1$. Conversely, if $\la_1$ and $\la_2$ are conjugates, then $SE_1$ and $SE_2$ are associated with the eigenvalues $\la_2 = \la_1^*$ and $\la_1 = \la_2^*$, respectively, so $F_1 = SE_1$ and $F_2 = SE_2$.

For the special case in which $L = R$, it is sufficient to consider the $2 \times 2$ rotation block and the associated $2 \times 1$ eigenvectors. $S$ can be omitted because its second and third components have the same sign. The eigenvalues of $R$ are $\la_+ = c + is$ and $\la_- = c - is = \la_+^*$,  the eigenvectors are $E_+ = [r, -ir]^t$ and $E_- = [r, ir]^t = E_+^*$ and the adjoint eigenvectors are $F_+ = E_+$ and $F_- = E_-$. The normalization conditions are $F_\pm^\d E_\pm =1$ and the orthogonality conditions are $F_\pm^\d E_\mp = 0$. Hence, $R$ has the adjoint decomposition $U\La U^\d$, where the columns of $U$ are the eigenvectors $E_+$ and $E_-$ (and $V = U$). These results are consistent with Eq. (\ref{3.7.9}).

The preceding discussions of the direct and adjoint eigenvalue problems involved the standard inner product $F^\d E$. But special relativity is based on the modified inner product $F^\d SE$, where $S$ is the metric matrix. (The squared interval is $T^tST$, where $T$ is the coordinate vector.) If one were to consider the weighted eigenvalue problems $LE = \la SE$ and $L^tF = \la^*SF$, one would find that $F_2^\d LE_1 = \la_1F_2^\d SE_1$, $E_1^\d L^tF_2 = \la_2^*E_1^\d SF_2$ and $(\la_2 - \la_1)F_2^\d SE_1 = 0$. The direct and adjoint eignvectors would be orthogonal relative to the modified inner product.

In three dimensions, it remains true that the eigenvalues occur in reciprocal pairs. It is helpful to consider some examples.
First, let $B_x$ represent a boost in the $x$ direction. Then $B_x$ has the eigenvalues $\ga + u$, $\ga - u$, 1 and 1, because the $y$ and $z$ axes are perpendicular to the boost axis. A similar statement can be made about a boost in an arbitrary direction, because rotation of the boost axis is a similarity transformation.
Second, let $R$ represent a rotation. Then $R$ is a block-diagonal matrix, in which the top-left component is 1 and the bottom-right block is $R_3$. The time-like vector $[1, 0, 0, 0]^t$ is an eigenvector of $R$, with eigenvalue 1. By combining the eigenvalue equation $R_3E = \la E$ with its hermitian conjugate, one finds that $|\la|^2E^\d E = E^\d R_3^tR_3E = E^\d E$. Hence, $|\la|^2 = 1$. $R_3$ has a pair of conjugate eigenvalues, which are reciprocals of each other, and a unit eigenvalue, which is its own reciprocal. The eigenvector associated with the unit eigenvalue is parallel to the rotation axis. In total, $R$ has the eigenvalues 1, 1, $c + is$ and $c - is$.
Third, let $L = R_xB_x$, where $R_x$ represents a rotation about the $x$ axis (in the $yz$ plane). Then $L$ is block diagonal, and has the eigenvalues $\ga + u$, $\ga - u$, $c + is$ and $c - is$. In three dimensions, the second pair of eigenvalues could be 1 and 1, but need not be.

\newpage

\subsec{3.8. Generators of Lorentz transformations}

Every Lorentz matrix $L$ (of practical interest) can be written as the exponential of a generating matrix $G$. Specifically, $L = \exp(Gk)$, where $k$ is a real parameter (Sec. 2). By differentiating the equation $L^tSL = S$ with respect to $k$ and taking the limit $k \rightarrow 0$, one obtains the generator equation
\be G^tS + SG = 0. \label{3.8.1} \ee
Three matrices that satisfy the generator equation are
\be G_1 = \left[\begin{array}{ccc} 0 & 1 & 0 \\ 1 & 0 & 0 \\ 0 & 0 & 0 \end{array}\right], \ \ 
G_2 = \left[\begin{array}{ccc} 0 & 0 & 1 \\ 0 & 0 & 0 \\ 1 & 0 & 0 \end{array}\right], \ \ 
G_3 = \left[\begin{array}{ccc} 0 & 0 & 0 \\ 0 & 0 & -1 \\ 0 & 1 & 0 \end{array}\right]. \label{3.8.2} \ee
Notice that these matrices all have zero trace. It follows from the identity $\det(L) = \exp[\tr(G)]$ that $\det(L) = 1$ if and only if $\tr(G) = 0$. Lorentz matrices have unit determinant (Sec. 3.1), so their generators must have zero trace.

To determine what transformations matrices (\ref{3.8.2}) generate, it is convenient to focus on their nontrivial $2 \times 2$ blocks. The upper-left block of $G_1$ has the property $(G_1')^2 = I$, where $I$ is the $2 \times 2$ identity matrix, from which it follows that
\ba \exp(G_1'k) &= &I + G_1'k + Ik^2/2 + G_1'k^3/6 + \dots \nonumber \\
&= &I\cosh(k) + G_1'\sinh(k). \label{3.8.3} \ea
The exponential of $G_2'$ is similar. The lower-right block of $G_3$ has the property $(G_3')^2 = -I$, from which it follows that
\ba \exp(G_3'k) &= &I + G_3'k - Ik^2/2 - G_3'k^3/6 + \dots \nonumber \\
&= &I\cos(k) + G_3'\sin(k). \label{3.8.4} \ea
By applying these results to the $3 \times 3$ generators, one obtains the fundamental transformation matrices
\be L_1 = \left[\begin{array}{ccc} C & S & 0 \\ S & C & 0 \\ 0 & 0 & 1 \end{array}\right], \ \ 
L_2 = \left[\begin{array}{ccc} C & 0 & S \\ 0 & 1 & 0 \\ S & 0 & C \end{array}\right], \ \ 
L_3 = \left[\begin{array}{ccc} 1 & 0 & 0 \\ 0 & c & -s \\ 0 & s & c \end{array}\right], \label{3.8.5}  \ee
where $C = \cosh(k)$, $S = \sinh(k)$, $c = \cos(k)$ and $s = \sin(k)$. The first and second matrices represent boosts in the $x$ and $y$ directions, respectively [Eqs. (\ref{3.2.1})], whereas the third represents a rotation about the $t$ axis (in the $xy$ plane) [Eq. (\ref{3.2.3})].

Notice that Eq. (\ref{3.8.1}) is linear in $G$. Hence, if $G_1$ -- $G_3$ satisfy the equation, so also does the linear combination $G_1k_1 + G_2k_2 + G_3k_3$, where the coefficients $k_1$ -- $k_3$ are arbitrary real numbers. The set of generating matrices is a vector space under addition \cite{taw91}, in which the matrices in Eq. (\ref{3.8.2}) play the roles of basis vectors.

One can exponentiate an arbitrary generating matrix by using the Cayley--Hamilton theorem \cite{gil08}. Let $\la_1$, $\la_2$ and $\la_3$ be the eigenvalues of $G$. Then $G$ satisfies the characteristic equation
\be G^3 - (\la_1 + \la_2 + \la_3)G^2 + (\la_1\la_2 + \la_2\la_3 + \la_3\la_1)G - \la_1\la_2\la_3I = 0, \label{3.8.11} \ee
from which it follows that
\be G^3 = \la_1\la_2\la_3I - (\la_1\la_2 + \la_2\la_3 + \la_3\la_1)G + (\la_1 + \la_2 + \la_3)G^2. \label{3.8.12} \ee
Hence, $\exp(G) = aI + bG + cG^2$, where $a$, $b$ and $c$ are functions of the eigenvalues.

Let $G = G_1k_1 + Gk_2 + G_3k_3$.
Then, written explicitly,
\be G = \left[\begin{array}{ccc} 0 & k_1 & k_2 \\ k_1 & 0 & -k_3 \\ k_2 & k_3 & 0 \end{array}\right]. \label{3.8.13} \ee
It is easy to verify that $G$ has the eigenvalues 0 and $\pm k$, where $k = (k_1^2 + k_2^2 - k_3^2)^{1/2}$. (In the rest of this section, $k_1$ -- $k_3$ are independent parameters, whereas $k$ is a dependent parameter.) It follows from Eq. (\ref{3.8.12}) that $G^3 = k^2G$. Hence, the exponential
\ba \exp(G) &= & I + G + G^2/2 + Gk^2/3! + G^2k^2/4! + Gk^4/5! + G^2k^4/6! \dots \nonumber \\
&= &I + G(1 + k^2/3! + k^4/5! \dots) + G^2(1/2 + k^2/4! + k^4/6! \dots) \nonumber \\
&= &I + G\sinh(k)/k + G^2[\cosh(k) - 1)]/k^2. \label{3.8.14} \ea
The squared matrix
\be G^2 = \left[\begin{array}{ccc} k_1^2 + k_2^2 & k_2k_3 & -k_1k_3 \\ -k_3k_2 & k_1^2 - k_3^2 & k_1k_2 \\ k_3k_1 & k_2k_1 & k_2^2 - k_3^2 \end{array}\right]. \label{3.8.15} \ee
By combining Eqs. (\ref{3.8.13}) -- (\ref{3.8.15}), one obtains the exponentiated matrix
\ba L &= &\left[\begin{array}{ccc} 1 + (n_1^2 + n_2^2)D & n_1S + n_2n_3D & n_2S - n_1n_3D \\
n_1S - n_3n_2D & 1 + (n_1^2 - n_3^2)D & -n_3S + n_1n_2D \\
n_2S + n_3n_1D & n_3S + n_2n_1D & 1 + (n_2^2 - n_3^2)D \end{array}\right] \nonumber \\
&= &\left[\begin{array}{ccc} C + n_3^2D & n_1S + n_2n_3D & n_2S - n_1n_3D \\
n_1S - n_3n_2D & C - n_2^2D & -n_3S + n_1n_2D \\
n_2S + n_3n_1D & n_3S + n_2n_1D & C - n_1^2D \end{array}\right], \label{3.8.16} \ea
where $C = \cosh(k)$, $D = C - 1$, $S = \sinh(k)$ and $n_i = k_i/k$.

First, consider the special case in which $k_3 = 0$ and $k = (k_1^2 + k_2^2)^{1/2}$. Then
\be L = \left[\begin{array}{ccc} C & n_1S & n_2S \\ n_1S & 1 + n_1^2D & n_1n_2D \\ n_2S & n_1n_2D & 1 + n_2^2D \end{array}\right]. \label{3.8.17} \ee
According to Eq. (\ref{3.3.2}), matrix (\ref{3.8.17}) decribes a boost with energy $\ga = C$ and momentum $u = S$ in the direction $(n_1, n_2) = (\cos\th, \sin\th)$. Conversely, if $\ga$ and $\th$ are specified, then $k =$ $\cosh^{-1}(\ga) = \log(\ga + u)$, $k_1 = k\cos\th$ and $k_2 = k\sin\th$. Notice that $\tr(L) = 2C + 1$ is positive.

Second, consider the complementary case in which $k_1 = k_2 = 0$ and $k = (-k_3^2)^{1/2} = ik_3$. Then
\be L = \left[\begin{array}{ccc} 1 & 0 & 0 \\ 0 & c_3  & -s_3 \\ 0 & s_3 & c_3 \end{array}\right], \label{3.8.18} \ee
where $c_3 = \cos(k_3)$ and $s_3 = \sin(k_3)$. Matrix (\ref{3.8.18}) describes a rotation through the angle $\th = k_3$, so there is no inverse problem to solve. Notice that $\tr(L) = 2c_3 + 1$ can be negative.

Third, consider the general case in which $n_1$, $n_2$, and $n_3$ are nonzero, and let $L = [l_{ij}]$. Then, by using the identities $n_1^2 + n_2^2 - n_3^2 = 1$ and $S^2 = D(D + 2)$, one can show that $l_{00}^2 - l_{01}^2 - l_{02}^2 = 1$ and $l_{00}^2 - l_{10}^2 - l_{20}^2 = 1$.
Hence, the first row of matrix (\ref{3.8.16}) can be written in the form $[\ga, uc_1, us_1]$ and the first column can be written in the form $[\ga, uc_2, us_2]^t$, where $\ga^2 - u^2 = 1$, $c_i = \cos(\th_i)$ and $s_i = \sin(\th_i)$, as stated in Eq. (\ref{3.3.12}). The energy and momentum~are
\be \ga = 1 + (n_1^2 + n_2^2)D, \ \ u =  [(n_1^2 + n_2^2)(S^2 + n_3^2D^2)]^{1/2}, \label{3.8.21} \ee
respectively, and the input and output angles are specified implicitly by the equations
\be \tan(\th_1) = {n_2S - n_1n_3D \over n_1S + n_2n_3D}, \ \ 
\tan(\th_2) = {n_2S + n_1n_3D \over n_1S - n_2n_3D}, \label{3.8.22} \ee
respectively. Notice that the only difference between these formulas is the sign of $n_3$, so the transformation is a boost ($\th_1 = \th_2$) if and only if $n_3 = 0$.
It follows from Eqs. (\ref{3.8.22}), and the trigonometric identities $c = 1/(1 + t^2)^{1/2}$ and $s = t/(1 + t^2)^{1/2}$, that
\ba c_1 = {n_1S + n_2n_3D \over [(n_1^2 + n_2^2)(S^2 + n_3^2D^2)]^{1/2}}, \ \
s_1 = {n_2S - n_1n_3D \over [(n_1^2 + n_2^2)(S^2 + n_3^2D^2)]^{1/2}}, \label{3.8.23} \\
c_2 = {n_1S - n_2n_3D \over [(n_1^2 + n_2^2)(S^2 + n_3^2D^2)]^{1/2}}, \ \
s_2 = {n_2S + n_1n_3D \over [(n_1^2 + n_2^2)(S^2 + n_3^2D^2)]^{1/2}}. \label{3.8.24} \ea
Notice that the denominator in these formulas is $u$, so the numerators are $uc_i$ and $us_i$.

It remains to be shown that the formulas for the components of the lower-right blocks of matrices (\ref{3.3.12}) and (\ref{3.8.16}) are equivalent. By combining Eqs. (\ref{3.8.22}), one finds that
\be \tan(\th_{21}) = {2n_3DS \over S^2 - n_3^2D^2}, \label{3.8.25} \ee
from which it follows that
\be c_{21} = {S^2 - n_3^2D^2 \over S^2 + n_3^2D^2}, \ \ 
s_{21} = {2n_3DS \over S^2 + n_3^2D^2}. \label{3.8.26} \ee
For the component $l_{11}$,
\ba c_{21} + \de c_2c_1 &= &{S^2 - n_3^2D^2 \over S^2 + n_3^2D^2}
+ (n_1^2 + n_2^2)D {(n_1S - n_2n_3D)(n_1S + n_2n_3D) \over [(n_1^2 + n_2^2)(S^2 + n_3^2D^2)]} \nonumber \\
&= &{S^2 - n_3^2D^2 \over S^2 + n_3^2D^2} + {(n_1^2S^2 - n_2^2n_3^2D^2)D \over S^2 + n_3^2D^2}. \label{3.8.27} \ea
The numerator in Eq. (\ref{3.8.27}) is
\ba &&S^2 - n_3^2D^2 + (n_1^2 + n_2^2)S^2D - (S^2 + n_3^2D^2)n_2^2D \nonumber \\
&= &S^2 - n_3^2D^2 + (1 + n_3^2)(D^2 + 2D)D - (S^2 + n_3^2D^2)n_2^2D \nonumber \\
&= &S^2 + n_3^2D^2 + D^3 + 2D^2 + n_3^2D^3 - (S^2 + n_3^2D^2)n_2^2D \nonumber \\
&= &S^2 + n_3^2D^2 + (S^2 + n_3^2D^2)D - (S^2 + n_3^2D^2)n_2^2D, \label{3.8.28} \ea
which is proportional to $C - n_2^2D$.
For the component $l_{21}$,
\ba s_{21} + \de s_2c_1 &= &{2n_3DS \over S^2 + n_3^2D^2}
+ (n_1^2 + n_2^2)D {(n_2S + n_1n_3D)(n_1S + n_2n_3D) \over [(n_1^2 + n_2^2)(S^2 + n_3^2D^2)]} \nonumber \\
&= &{2n_3DS \over S^2 + n_3^2D^2} + {[n_1n_2S^2 + (n_1^2 + n_2^2)n_3DS + n_1n_2n_3^2D^2]D \over S^2 + n_3^2D^2}. \label{3.8.29} \ea
The numerator in Eq. (\ref{3.8.29}) is
\ba &&[2D + (1 + n_3^2)D^2]n_3S + (S^2 + n_3^2D^2)n_1n_2D \nonumber \\
&= &[D(D + 2) + n_3^2D^2]n_3S + (S^2 + n_3^2D^2)n_1n_2D \nonumber \\
&= &(S^2 + n_3^2D^2)n_3S + (S^2 + n_3^2D^2)n_1n_2D, \label{3.8.30} \ea
which is proportional to $n_3S + n_1n_2D$. [In Eqs. (\ref{3.8.28}) and (\ref{3.8.30}), the identities mentioned before Eq. (\ref{3.8.21}) were used repeatedly.] The proofs of the equivalences of the formulas for $l_{12}$ and $l_{22}$ are similar.
Not only does the preceding analysis show that Eqs. (\ref{3.3.12}) and (\ref{3.8.16}) are equivalent, but it is also a constructive proof of the former equation. (An elegant, but abstract, proof was provided in Sec. 3.5.)

Conversely, suppose that $\ga$, $\th_1$ and $\th_2$ are specified. Then, by comparing the traces of matrices (\ref{3.3.12}) and (\ref{3.8.16}), one finds that
\be C = [\ga - 1 + (\ga + 1)c_{21}]/2. \label{3.8.31} \ee
With $C$ known, so also are $D = C - 1$, $S = (C^2 - 1)^{1/2}$ and $k = \log(C + S)$. By adding and subtracting pairs of the off-diagonal components, one finds that
\be n_1 = u(c_2 + c_1)/2S, \ \ n_2 \ = \ u(s_2 + s_1)/2S, \ \ n_3 \ = \ (\ga + 1)s_{21}/2S. \label{3.8.32} \ee
In Eqs. (\ref{3.8.16}) and (\ref{3.8.32}), changing the sign of $S$ is equivalent to changing the signs of $n_i$, so one can assume that $S \ge 0$ without loss of generality.

It follows from the definition $L = \exp(G)$ that the inverse matrix $L^{-1} = \exp(-G)$, where, according to Eq. (\ref{3.8.14}),
\be \exp(-G) = I - GS/k + G^2D/k^2. \label{3.8.41} \ee
By combining Eqs. (\ref{3.8.16}) and (\ref{3.8.41}), one finds that
\ba e^{-G}e^G &= &(I - GS/k + G^2D/k^2)(I + GS/k + G^2D/k^2) \nonumber \\
&= &(I + G^2D/k^2)^2 - G^2S^2/k^2 \nonumber \\
&= &I + 2G^2D/k^2 + G^4D^2/k^4 - G^2S^2/k^2. \label{3.8.42} \ea
By using the identity $G^3 = k^2G$, one can rewrite the right side of Eq. (\ref{3.8.42}) as the sum of $I$ and a term that is proportional to $G^2/k^2$. The coefficient of this term is
\ba 2D + D^2 - S^2 = 2(C - 1) + (C^2 - 2C + 1) - S^2 = 0. \label{3.8.43} \ea
Equation (\ref{3.8.43}) confirms that the inverse of $\exp(G)$ is $\exp(-G)$.
Changing the sign of $G$ is equivalent to changing the sign of $S$, or the signs of $n_i$. Hence, the inverse matrix
\ba L^{-1} &= &\left[\begin{array}{ccc} C + n_3^2D & -n_1S + n_2n_3D & -n_2S - n_1n_3D \\
-n_1S - n_3n_2D & C - n_2^2D & n_3S + n_1n_2D \\
-n_2S + n_3n_1D & -n_3S + n_2n_1D & C - n_1^2D \end{array}\right]. \label{3.8.44} \ea
By comparing Eqs. (\ref{3.8.16}) and (\ref{3.8.44}), one finds that $L^{-1} = SL^tS$, where $S$ is the metric matrix,  as stated in Sec. 3.1. It follows from eqs. (\ref{3.8.17}), (\ref{3.8.18}) and (\ref{3.8.44}) that the inverse of a boost is a boost in the opposite direction, and the inverse of a rotation is the opposite rotation (which has the same angle, but the opposite sense).

The derivation of formula (\ref{3.8.16}) was based on the assumption that $k_1^2 + k_2^2 - k_3^2 > 0$. In the opposite case, $k \rightarrow ik = i(k_3^2 - k_1^2 - k_2^2)^{1/2}$, $\sinh(k)/k \rightarrow \sin(k)/k$ and $[\cosh(k) - 1]/k^2 \rightarrow [1 - \cos(k)]/k^2$. With these changes, formulas (\ref{3.8.16}) and (\ref{3.8.44}) remain valid. ($G^3 \rightarrow -k^2G$, because the definition of $k$ changes.) In both cases, $\ga = l_{00} > 1$. Boosting coexists with rotation.

The generator method is useful, because it shows how complicated transformations are related to simple ones (boosts and rotations). It also allows one to identify hidden relations between different matrix groups. It is easy to verify that generators (\ref{3.8.2}) satisfy the commutation relations
\be [G_1, G_2] = -G_3, \ \ [G_2, G_3] = G_1, \ \ [G_3, G_1] = G_2, \label{3.8.51} \ee
where the commutator $[x, y] = xy - yx$. The generators of the symplectic group Sp(2), which consists of real $2 \times 2$ matrices, and the special unitary group SU(1,1), which consists of complex $2 \times 2$ matrices, satisfy the same relations \cite{kim83,mck25a,mck25c}.

Consider two matrix groups, whose members have the same number of free parameters. Let $A$, $B$ and $C$ be members of group one, $D$, $E$ and $F$ be members of group two, and suppose that there is a one-to-one relation between the groups. Then every object matrix $A$ has one image $D$, and every image matrix $E$ has one object $B$, of which it is the image. This relation between the groups is called an isomorphism if it preserves multiplication: $BA = C$ in group one if and only if $ED = F$ in group two, where $F$ is the image of $C$. If two groups are isomorphic, they have the same structure. For example, if $B = A^{-1}$, then $E = D^{-1}$. If $A$, $B$ and $C$ form a subgroup of group one, then $D$, $E$ and $F$ form a subgroup of group two.

%Suppose that $A = \exp(G_1k_1 + G_2k_2 + G_3k_3)$ and $B = \exp(G_1l_1 + G_2l_2 + G_3l_3)$. Then, by using the commutation relations (of group one), one can write $A$ and $B$ in normal form, in which (for example) the powers of $G_3$ appear to the left of (before) those of $G_2$, which appear before those of $G_1$. (An example of normal ordering is provided in App. D.) Although the product $BA$ is not necessarily in normal form, one can put it in normal form by using the commutation relations. Now suppose that $D = \exp(H_1k_1 + H_2k_2 + H_3k_3)$ and $E = \exp(H_1l_1 + H_2l_2 + H_3l_3)$.
Suppose that $A = \exp(\tsum_i G_i k_i)$ and $B = \exp(\tsum_i G_i l_i)$. Then, by using the commutation relations (of group one), one can write $A$ and $B$ in normal form, in which (for example) the powers of $G_3$ appear to the left of (before) those of $G_2$, which appear before those of $G_1$. (An example of normal ordering is provided in App. D.) Although the product $BA$ is not necessarily in normal form, one can put it in normal form by using the commutation relations. Now suppose that $D = \exp(\tsum_i H_i k_i)$ and $E = \exp(\tsum_i H_i l_i)$.
[The associated matrices have the same coefficients ($k_i$ and $l_i$), but different generators ($G_i$ and $H_i$).] Then, by using the commutation relations (of group two), one can write $D$, $E$ and $ED$ in normal form. If the commutation relations of the two groups are the same, then so also will be the normal forms of the product matrices $BA$ and $ED$. Hence, the rules of multiplication are preserved, so the groups are isomorphic.
If two groups are isomorphic, one can establish results for the easier-to-analyze group (for example, $2 \times 2$ matrices) and know, without further effort, that they are also true for the harder-to-analyze group (for example, $3 \times 3$ matrices).

Even if some image matrices in group two have more than one object matrix in group one, it might still be true that $C = BA$ implies that $F = ED$, in which case the relation between groups one and two is called a homomorphism. (The inverse relation is ambiguous.) The homomorphisms between Sp(2) or SU(1,1) and SO(1,2) are discussed in detail in \cite{mck25c}.

In three dimensions, every Lorentz matrix has six generators, three of which represent boosts along the $x$, $y$ and $z$ axes, and three of which represent rotations about the $x$, $y$ and $z$ axes \cite{jac99}.

\newpage

\subsec{3.9. Combinations of Lorentz transformations}

In Sec. 3.4, we discussed briefly the combination of two arbitrary boosts. The complexity of the algebra obscured the physics of the combined transformation. Fortunately, the theory developed in Sec. 3.6 (Schmidt-like decompositions) allows the combination problem to be solved completely and its solution to be interpreted easily.

First, consider a simple example that involves perpendicular boosts. Let $L_1$ and $L_2$ be matrices that describe boosts in the $x$ and $y$ directions, respectively. (One can always align the $x$ axis with the direction of the first boost, so there is no loss of generality in considering boosts in these directions.) Then %the product matrix
\ba L_2L_1 &= &\left[\begin{array}{ccc} \ga_2 & 0 & u_2 \\ 0 & 1 & 0 \\ u_2 & 0 & \ga_2 \end{array}\right]
\left[\begin{array}{ccc} \ga_1 & u_1 & 0 \\ u_1 & \ga_1 & 0 \\ 0 & 0 & 1 \end{array}\right] \nonumber \\
&= &\left[\begin{array}{ccc} \ga_2\ga_1 & \ga_2u_1 & u_2 \\ u_1 & \ga_1 & 0 \\ u_2\ga_1 & u_2u_1 & \ga_2 \end{array}\right], \label{3.9.1} \ea
where $\ga_i$ and $u_i$ are the energy and momentum parameters, respectively.
The product matrix
$L_{21} = L_2L_1$ is a Lorentz matrix, but it is asymmetric, so it does not represent a boost. (We will show shortly that it is a boost followed by a rotation.) If the boosts are made in the opposite order, then
\ba L_1L_2 &= &\left[\begin{array}{ccc} \ga_1 & u_1 & 0 \\ u_1 & \ga_1 & 0 \\ 0 & 0 & 1 \end{array}\right]\left[\begin{array}{ccc} \ga_2 & 0 & u_2 \\ 0 & 1 & 0 \\ u_2 & 0 & \ga_2 \end{array}\right] \nonumber \\
&= &\left[\begin{array}{ccc} \ga_1\ga_2 & u_1 & \ga_1u_2 \\ u_1\ga_2 & \ga_1 & u_1u_2 \\ u_2 & 0 & \ga_2 \end{array}\right]. \label{3.9.2} \ea
Notice that $L_1L_2 = L_1^tL_2^t = (L_2L_1)^t$, because boost matrices are symmetric. This example shows that boost matrices do not necessarily commute. Nonetheless, the energy $\ga_2\ga_1$ is independent of the order of the boosts. (In Sec. 3.4, we showed that this statement is true for arbitrary boosts.)

In Sec. 3.5, we showed that every Lorentz matrix can be written in the form of Eq. (\ref{3.3.12}), where $\ga$ and $u = (\ga^2 - 1)^{1/2}$ are the energy and momentum parameters, respectively, $\de = \ga - 1$, $\th_1$ and $\th_2$ are the input and output angles, respectively, and $\th_{21} = \th_2 - \th_1$ is the difference angle.
Matrix (\ref{3.9.1}) can be written in this form, so the energy $\ga = \ga_2\ga_1$. The input and output angles are specified implicitly by the equations
\be \tan(\th_1) = u_2/\ga_2u_1, \ \ \tan(\th_2) = u_2\ga_1/u_1,  \label{3.9.4} \ee
respectively. By using trigonometric identities, one finds that
\ba c_1 &= &{\ga_2u_1/(\ga_2^2u_1^2 + u_2^2)^{1/2}}, \ \ s_1 \ = \ {u_2/(\ga_2^2u_1^2 + u_2^2)^{1/2}}, \label{3.9.5} \\
c_2 &= &{u_1/(u_2^2\ga_1^2 + u_1^2)^{1/2}}, \ \ s_2 \ = \ {u_2\ga_1/(u_2^2\ga_1^2 + u_1^2)^{1/2}}. \label{3.9.6} \ea
It is easy to show that both denominators equal $(\ga_2^2\ga_1^2 - 1)^{1/2} = [(\ga_2\ga_1 + 1)(\ga_2\ga_1 - 1)]^{1/2}$, from which it follows that
\ba c_{21} &= &{\ga_2u_1^2 + u_2^2\ga_1 \over \ga_2^2\ga_1^2 - 1} 
\ = \ {(\ga_2 + \ga_1)(\ga_2\ga_1 - 1) \over (\ga_2\ga_1 + 1)(\ga_2\ga_1 - 1)}
\ = \ {\ga_2 + \ga_1 \over \ga_2\ga_1 + 1}, \label{3.9.7} \\
s_{21} &= &{(u_2\ga_1)(\ga_2u_1) - u_1u_2 \over \ga_2^2\ga_1^2 - 1}
\ = \ {u_2u_1(\ga_2\ga_1 - 1) \over (\ga_2\ga_1 + 1)(\ga_2\ga_1 - 1)}
\ = \ {u_2u_1 \over \ga_2\ga_1 + 1}. \label{3.9.8} \ea
By combining Eqs. (\ref{3.9.7}) and (\ref{3.9.8}), one finds that the difference angle is specified by the equation
\be \tan(\th_{21}) = u_2u_1/(\ga_2 + \ga_1). \label{3.9.9} \ee
Notice that Eq. (\ref{3.9.9}) is consistent with Eqs. (\ref{3.5.14}) and (\ref{3.9.1}).

It remains to evaluate the terms in the lower-right block of matrix (\ref{3.3.12}). Terms that are quadratic in $c_i$ or $s_i$ have denominators of $\ga_2^2\ga_1^2 - 1$ and $\de = \ga_2\ga_1 - 1$, so the products of such terms have denominators of $\ga_2\ga_1 + 1$, as do the denominators of $c_{21}$ and $s_{21}$. Hence,
\ba c_{21} + \de c_2c_1 &\propto &\ga_2 + \ga_1 + \ga_2(\ga_1^2 - 1)
\ = \ \ga_1(\ga_2\ga_1 + 1), \label{3.9.10} \\
-s_{21} + \de c_2s_1 &\propto &-u_2u_1 + u_1u_2 \ = 0, \label{3.9.11} \\
 s_{21} + \de s_2c_1 &\propto &u_2u_1 + (u_2\ga_1)(\ga_2u_1)
\ = \ u_2u_1(\ga_2\ga_1 + 1), \label{3.9.12} \\
 c_{21} + \de s_2s_1 &\propto &\ga_2 + \ga_1 + (\ga_2^2 - 1)\ga_1
\ = \ \ga_2(\ga_2\ga_1 + 1). \label{3.9.13} \ea
By dividing Eqs. (\ref{3.9.10}) -- (\ref{3.9.13}) by $\ga_2\ga_1 + 1$, one obtains the entries of the lower-right block of matrix (\ref{3.9.1}). Thus, $L_{21}$ has the Schmidt-like decomposition $R_2BR_1^t = R_{21}(R_1BR_1^t)$, where $B$ is a boost matrix, and $R_i = R(\th_i)$ is a rotation matrix. The composition of two perpendicular boosts is a boost whose axis is inclined at the angle $\th_1$ relative to the $x$ axis, followed by a rotation through the angle $\th_{21}$.
In the context of boosts, this rotation angle is called the Wigner angle.

The inverse of two boosts is two brakes, done in the opposite order. Hence,
\ba (L_2L_1)^{-1} &= &\left[\begin{array}{ccc} \ga_1 & -u_1 & 0 \\ -u_1 & \ga_1 & 0 \\ 0 & 0 & 1 \end{array}\right]\left[\begin{array}{ccc} \ga_2 & 0 & -u_2 \\ 0 & 1 & 0 \\ -u_2 & 0 & \ga_2 \end{array}\right] \nonumber \\
&= &\left[\begin{array}{ccc} \ga_1\ga_2 & -u_1 & -\ga_1u_2 \\ -u_1\ga_2 & \ga_1 & u_1u_2 \\ -u_2 & 0 & \ga_2 \end{array}\right]. \label{3.9.14} \ea
By comparing Eqs. (\ref{3.9.2}) and (\ref{3.9.14}), one finds that $(L_2L_1)^{-1} = S(L_2L_1)^tS$, where $S$ is the metric matrix, as stated in Sec. 3.1.

Second, let $L_1$ and $L_2$ be arbitrary transformation matrices. Then each matrix has the Schmidt-like decomposition $QBP^t$, where $B(\ga)$ represents a boost in the $x$ direction, and $P(\ph)$ and $Q(\th)$ represent two-dimensional rotations. (In the current context, it is better to use different symbols for the input and output rotation matrices and angles.) Hence, the product matrix
\be L_2L_1 = Q_2B_2P_2^tQ_1B_1P_1^t = Q_2(B_2R_{21}^tB_1)P_1^t, \label{3.9.21} \ee
where $R_{21}^t = P_2^tQ_1 = Q_1P_2^t = (P_2Q_1^t)^t$ is also a rotation matrix.
(In two dimensions, rotation matrices commute.)
The key product is the intermediate matrix $L_3 = B_2R_{21}^tB_1$.
Written explicitly,
\ba L_3 &= &\left[\begin{array}{ccc} \ga_2 & u_2 & 0 \\ u_2 & \ga_2 & 0 \\ 0 & 0 & 1 \end{array}\right]
\left[\begin{array}{ccc} 1 & 0 & 0 \\ 0 & c & s \\ 0 & -s & c \end{array}\right]
\left[\begin{array}{ccc} \ga_1 & u_1 & 0 \\ u_1 & \ga_1 & 0 \\ 0 & 0 & 1 \end{array}\right] \nonumber \\
&= &\left[\begin{array}{ccc} \ga_2 & u_2 & 0 \\ u_2 & \ga_2 & 0 \\ 0 & 0 & 1 \end{array}\right]
\left[\begin{array}{ccc} \ga_1 & u_1 & 0 \\ cu_1 & c\ga_1 & s \\ -su_1 & -s\ga_1 & c \end{array}\right] \nonumber \\
&= &\left[\begin{array}{ccc} \ga_2\ga_1 + u_2u_1c & \ga_2u_1 + u_2\ga_1c & u_2s \\ u_2\ga_1 + \ga_2u_1c & u_2u_1 + \ga_2\ga_1c & \ga_2s \\ -u_1s & -\ga_1s & c\end{array}\right], \label{3.9.22} \ea
where $c = \cos(\ph_2 - \th_1)$ and $s = \sin(\ph_2 - \th_1)$. This matrix is the product of Lorentz matrices, so it is also a Lorentz matrix, with the decomposition $L_3 = Q_3(\th_3)B_3(\ga_3)P_3^t(\ph_3)$. According to Eq. (\ref{3.3.12}), the first row of the product matrix is $[\ga_3, u_3c_\ph, u_3s_\ph]$, where $c_\ph = \cos(\ph_3)$, and the first column is $[\ga_3, u_3c_\th, u_3s_\th]^t$, where $c_\th = \cos(\th_3)$. The definitions of $s_\ph$ and $s_\th$ are similar.
It follows from Eqs. (\ref{3.3.12}) and (\ref{3.9.22}) that the intermediate energy
\be \ga_3 = \ga_2\ga_1 + u_2u_1c. \label{3.9.23} \ee
It also follows that the intermediate angles are specified by the equations
\be \tan(\ph_3) = {u_2s \over \ga_2u_1 + u_2\ga_1c}, \ \ 
\tan(\th_3) = {-u_1s \over u_2\ga_1 + \ga_2u_1c}. \label{3.9.24} \ee
Equations (\ref{3.9.23}) and (\ref{3.9.24}) are the product rules for Lorentz matrices, written in terms of the Schmidt-like parameters $\ga$, $\ph$ and $\th$. Notice that the intermediate energy is a symmetric function of $\ga_1$ and $\ga_2$, and depends on the difference angle $\ph_2 - \th_1$. (If the order of the transformations were reversed, it would depend on the angle $\ph_1 - \th_2$.) Given the complexity of matrices of the form (\ref{3.3.12}), it is remarkable that the product rules are so simple.
Now let $L_4 = L_2L_1$. Then $L_4 = Q_2L_3P_1^t$, from which it follows that
\be \ga_4 = \ga_3, \ \ \ph_4 = \ph_1 + \ph_3, \ \ \th_4 = \th_2 + \th_3. \label{3.9.25} \ee
Equations (\ref{3.9.23}) -- (\ref{3.9.25}) are valid for arbitrary transformations (combinations of boosts and rotations).

For perpendicular boosts, $\ph_1 = \th_1 = 0$ and $\ph_2 = \th_2 = \pi/2$, so in Eqs. (\ref{3.9.23}) and (\ref{3.9.24}), $c = 0$ and $s = 1$. [See the comment before Eq. (\ref{3.9.1}).]
The energy $\ga_4 = \ga_2\ga_1$ and the input angle is specified by the equation $\tan(\ph_4) = u_2/\ga_2u_1$, as stated in the first of Eqs. (\ref{3.9.4}), respectively. By using the identity $\tan(\th_3 + \pi/2) = -1/\tan(\th_3)$, one finds that the output angle is specified by the equation $\tan(\th_4) = u_2\ga_1/u_1$, as stated in the second of Eqs. (\ref{3.9.4}).
By combining these results, one obtains Eq. (\ref{3.9.9}), which specifies the Wigner angle $\th_4 - \ph_4$.

For arbitrary boosts, $\ph_1 = \th_1 = 0$ and $\ph_2 = \th_2 \neq 0$.  Hence, the product matrix
\ba B_2B_1 &= &\left[\begin{array}{ccc} \ga_2 & u_2c & u_2s \\ u_2c & 1 + \de_2c^2 & \de_2cs \\
u_2s & \de_2cs & 1 + \de_2s^2 \end{array}\right]
\left[\begin{array}{ccc} \ga_1 & u_1 & 0 \\u_1 & \ga_1 &0 \\
0 & 0 & 1 \end{array}\right] \nonumber \\
&= &\left[\begin{array}{ccc} \ga_2\ga_1 + u_2u_1c & \ga_2u_1 + u_2\ga_1c & u_2 s \\
(u_2\ga_1 + \de_2u_1c)c + u_1 & (u_2u_1 + \de_2\ga_1c)c + \ga_1 & \de_2cs \\
(u_2\ga_1 + \de_2u_1c)s & (u_2u_1 + \de_2\ga_1c)s & 1 + \de_2s^2 \end{array}\right], \label{3.9.26} \ea
where $c = c_{21}$ and $s = s_{21}$. (In a frame that is aligned with the first boost, the second boost angle is the difference angle $\th_{21}$.) The top-left component of matrix (\ref{3.9.26}) is the boost energy (\ref{3.9.23}). It follows from Eqs. (\ref{3.5.13}) and (\ref{3.9.26}) that
\be \tan(\ph_4) = {u_2s \over \ga_2u_1 + u_2\ga_1c}, \ \ 
\tan(\th_4) = {(u_2\ga_1 + \de_2u_1c)s \over (u_2\ga_1 + \de_2u_1c)c + u_1}. \label{3.9.27} \ee
Equations (\ref{3.9.27}) are consistent with Eqs. (\ref{3.9.24}) and (\ref{3.9.25}). %, in the second of which $\ph_1 = \th_4 = 0$ and $\th_2 = \th_{21}$.
It follows from Eqs. (\ref{3.5.14}) and (\ref{3.9.26}) that the Wigner angle is specified by the equation
\be \tan(\th_4 - \ph_4) = {(u_2u_1 + \de_2\de_1c)s \over \ga_2 + \ga_1 + (u_2u_1 + \de_2\de_1c)c}. \label{3.9.28} \ee
Notice that this angle is independent of the first boost angle, because $(\th_4 + \th_1) - (\ph_4 + \th_1) = \th_4 - \ph_4$.
One can derive an equivalent equation for the Wigner angle by using spinors \cite{van84,tor23a,tor23b}, which are analogs of Jones vectors in polarization optics \cite{gor00}.

In App. B, we use the standard vector formalism to show that the energy of a composite (two-stage) boost is
\be \ga = \ga_2\ga_1 + \vu_2\cdot\vu_1, \label{3.9.31} \ee
and the momentum row- and column-vectors are
\be \vu_2 + (\ga_2 + \ep_1\vu_2\cdot\vu_1)\vu_1, \ \ 
\vu_1 + (\ga_1 + \ep_2\vu_2\cdot\vu_1)\vu_2, \label{3.9.32} \ee
respectively, where $\ep_i = 1/(\ga_i + 1)$. The $x$ and $y$ components of the row vector are
\be (\ga_2u_1 + u_2\de_1c)c_1  + u_2c_2, \ \ 
(\ga_2u_1 + u_2\de_1c)s_1  + u_2s_2, \label{3.9.33} \ee
respectively, and the components of the column vector are
\be (u_2\ga_1 + \de_2u_1c)c_2  + u_1c_1, \ \ 
(u_2\ga_1 + \de_2u_1c)s_2  + u_1s_1. \label{3.9.34} \ee
Equation (\ref{3.9.31}) is identical to Eq. (\ref{3.9.23}). By making the substitutions $c_1 = 1$, $s_1 = 0$, $c_2 = c$ and $s_2 = s$ in Eqs. (\ref{3.9.33}) and (\ref{3.9.34}), one obtains formulas for $\tan(\ph_4)$ and $\tan(\th_4)$ that are consistent with the first and second of Eqs. (\ref{3.9.27}), respectively. Conversely, by using Eqs. (\ref{3.9.27}) to calculate $\tan(\ph_4 + \th_1)$ and $\tan(\th_4 + \th_1)$, one obtains formulas that are consistent with Eqs. (\ref{3.9.33}) and (\ref{3.9.34}). 
Thus, the matrix formalism (Secs. 3.5 and 3.6) is consistent with the vector formalism (App. B). It is also more general, because it applies to arbitrary transformations, not just boosts.

\newpage

\sec{4. Summary}

In this article, we reviewed the theory of Lorentz transformations in time and two space dimensions. This subject is a nontrivial generalization of the theory of transformations in one dimension, which includes most of the complexity and richness of transformations in three dimensions.

To introduce the basic concepts of Lorentz transformations, in Sec. 2 we discussed transformations in time and one space dimension. These transformations can be written in the matrix form $T' = LT$, where $L$ is a real $2 \times 2$ matrix, and $T = [t, x]^t$ and $T' = [t', x']^t$ are $2 \times 1$ input and output coordinate vectors, respectively [Eqs. (\ref{2.11}) and (\ref{2.12}]. The Lorentz matrix $L$ is specified by two (dimensionless) parameters, the momentum $u$ and energy $\ga = (1 + u^2)^{1/2}$, only one of which is independent (free). Lorentz transformations preserve the squared interval $t^2 - x^2 = (t')^2 - (x')^2$. The composition of two Lorentz transformations is also a transformation, and the composite transformation matrix is the product of the constituent transformation matrices.
%Standard formulas, for the addition and subtraction of velocities, are derived and discussed briefly.
The set of Lorentz matrices forms a group under multiplication, which is called the special orthogonal group SO(1,1). Every Lorentz matrix can be written as the exponential of a generating matrix: $L = \exp(Gk)$, where $k$ is a free parameter, to which $u$ is simply related [Eq. (\ref{2.23})].

Now let $L$ be a real $3 \times 3$ matrix, $T = [t, x, y]^t$ be a coordinate vector and $S = \diag(1,-1,-1)$ be the metric matrix. Then $L$ is a Lorentz matrix if and only if it satisfies the Lorentz condition $L^tSL = S$ [Eq. (\ref{3.1.1})], which guarantees the invariance of the squared interval $T^tST = t^2 - x^2 - y^2$. (If $U = [\ga, u_x, u_y]^t$ is an energy--momentum vector, then the same condition guarantees the invariance of $U^tSU = \ga^2 - u_x^2 - u_y^2$.) In Sec. 3.1, we showed that the set of Lorentz matrices forms a group under multiplication. This group is called the special orthogonal group SO(1,2) and its members are specified by three free parameters.

Let $B_x$ and $B_y$ be matrices that represent boosts in the $x$ and $y$ directions, which preserve the squared intervals $t^2 - x^2$ and $t^2 - y^2$, respectively, and let $R$ represent a rotation about the $t$ axis (in the $xy$ plane), which preserves the squared distance $x^2 + y^2$. Then, in Sec. 3.2, we showed that $B_x$, $B_y$ and $R$ are Lorentz matrices. So also are arbitrary products of these fundamental matrices. Boost matrices are symmetric, whereas rotation matrices are asymmetric [Eqs. (\ref{3.2.1}) and (\ref{3.2.3})].

In Sec. 3.3, we discussed combinations of rotations, and boosts and rotations. The product of two rotation matrices is another rotation matrix and in two dimensions, rotation matrices commute [Eq. (\ref{3.3.1})].
A matrix $B(\ga, \th)$ that describes a boost in an aritrary direction can be written as the product $RB_xR^t$, where $B_x(\ga)$ represents a boost in the $x$ direction with energy $\ga$ and $R(\th)$ represents a rotation through the angle $\th$ [Eq. (\ref{3.3.2})].

In Sec. 3.4, we discussed briefly the combination of two boosts $B_1$ and $B_2$. The product matrix $B_2B_1$ is not necessarily symmetric, so the composite transformation is not necessarily a boost. The formulas for the components of the product matrix are complicated and unilluminating [Eqs. (\ref{3.4.2}) -- (\ref{3.4.10})], so further work was required.

In Sec. 3.5, we used the Lorentz condition to derive the general form of a Lorentz matrix, which is boost matrix followed, or preceded, by a rotation matrix [Eq. (\ref{3.5.11})].
Let $\ga$, $\th_1$ and $\th_2$ be arbitrary parameters, in which case $\th_{21} = \th_2 - \th_1$ is also arbitrary. Then $L = R_{21}(R_1B_xR_1^t) = R_2B_xR_1^t = (R_2B_xR_2^t)R_{21}$, where $R_i = R(\th_i)$ and $R_{21} = R_2R_1^t$. Notice that a boost followed by a rotation is equivalent to the same rotation followed by a boost.
Although the boost matrices differ, they represent the same boost, expressed in different coordinate systems.
The natural set of free parameters consists of the boost energy and two rotation angles [Eqs. (\ref{3.5.12}) -- (\ref{3.5.14})].

Every real matrix $M$ has the Schmidt decomposition $M = QDP^t$, where $D$ is a nonnegative diagonal matrix, and $P$ and $Q$ are orthogonal matrices. The components of $D$ are called Schmidt coefficients and specify dilations. The columns of $P$ and $Q$ are called input and output Schmidt vectors, respectively. Both sets of vectors are orthogonal.
In Sec. 3.6, we used matrix (\ref{3.5.11}) to derive the Schmidt decomposition of a Lorentz matrix [Eq. (\ref{3.6.4})]. The Schmidt coefficients are $\ga + u$, $\ga - u$ and 1 [Eqs. ({\ref{3.6.7}) and (\ref{3.6.10})], which correspond to stretching, squeezing and conservation, respectively. Notice that the first and second coefficients are reciprocals. The associated input vectors define the characteristic coordinates $t + \xi$, $t - \xi$ and $\et$, where $\xi $ and $\et$ represent distance parallel and perpendicular to the boost axis, respectively [Eqs. (\ref{3.6.6})]. The output vectors are rotated versions of the input vectors (because a Lorentz matrix represents a boost followed by a rotation) [Eqs. (\ref{3.6.9})].
By using  Eq. (\ref{3.6.11}), we showed that every Lorentz matrix has the Schmidt-like decomposition $L = QBP^t$, where $B$ represents a boost (rather than a dilation), and $P$ and $Q$ represent two-dimensional rotations (rather than three-dimensional orthogonal transformations) [Eqs. (\ref{3.6.12}) -- (\ref{3.6.14})]. In the notation of the preceding paragraph, $L = R_2B_xR_1^t$.

In Sec. 3.7, we used the Schmidt-like decomposition of a Lorentz matrix to determine its spectral decomposition. The eigenvalues of $L$ can be real or complex, depending on whether the transformation is boost-like or rotation-like [Eq. (\ref{3.7.1})]. One eigenvalue is always 1, and the other two eigenvalues are reciprocals. If these eigenvalues are real, then they are greater and less than 1, whereas if they are complex, they are conjugates of modulus 1. Real eigenvalues correspond to dilations (a stretch and a squeeze) in the directions of the associated real eigenvectors, which are not necessarily orthogonal, whereas complex eigenvalues correspond to rotation about an axis that is specified by the real and imaginary parts of the associated complex eigenvectors, which are also not necessarily orthogonal.

Every Lorentz matrix (of practical interest) can be written as the exponential of a generating matrix $G$. In two dimensions, there are three basis generators $G_i$, of which $G = \tsum_i G_ik_i$ is a linear combination. %$G = G_1k_1 + G_2k_2 + G_3k_3$
$G_1$ and $G_2$ produce boosts in the $x$ and $y$ directions, respectively, whereas $G_3$ produces a rotation about the $t$ axis [Eqs. (\ref{3.8.2}) and (\ref{3.8.5})].
In Sec. 3.8, we used the Cayley--Hamilton theorem to exponentiate the generating matrix efficiently. By doing so, we rederived the general form of a Lorentz matrix from first principles [Eq. (\ref{3.8.16})]. Every Lorentz matrix is specified by three parameters: The boost energy $\ga$, and the rotation angles $\th_1$ and $\th_2$, or alternatively, the generator coefficients $k_1$, $k_2$ and $k_3$. We derived formulas for the boost and angle parameters in terms of the generator coefficients [Eqs. (\ref{3.8.21}) and (\ref{3.8.22})], and inverse formulas for the generator coefficients in terms of the boost and angle parameters [Eqs. (\ref{3.8.31}) and (\ref{3.8.32})].

The symplectic group Sp(2) and the special unitary group SU(1,1) consist of real and complex $2 \times 2$ matrices, respectively. The generators of these groups differ from those of SO(1,2), but they satisfy the same commutation relations [Eqs. (\ref{3.8.51})]. Hence, Sp(2), SU(1,1) and SO(1,2) are homomorphic (have similar structures). One can use this fact to deduce properties of SO(1,2) from known properties of Sp(2) and SU(1,1) \cite{kim83,mck25a,mck25c,tor23b}.

In Sec. 3.9, we discussed in detail the combination of two arbitrary transformations. By using Schmidt-like decompositions of the constituent matrices $L_1$ and $L_2$, we derived formulas for the boost and angle parameters of the product matrix $L_2L_1$ [Eqs. (\ref{3.9.23}) -- (\ref{3.9.25})], which, like any other Lorentz matrix, can be written as a boost matrix followed by a rotation matrix. Despite the complexity of the constituent matrices [Eq. (\ref{3.3.12})], the product formulas are relatively simple. For the special case in which the constituent transformations are boosts [Eq. (\ref{3.3.2}) or (\ref{3.3.11})], the formulas derived using the matrix formalism reduce to those derived using the vector formalism (App. B). In particular, the formulas for the rotation (Wigner) angle are equivalent [Eqs. (\ref{3.9.28}) and (\ref{b51})].

In summary, despite the mathematical complexities of Lorentz transformations in time and two space dimensions, the transformations can be understood in relatively simple physical terms. Many of the results derived herein for two dimensions can be generalized to three dimensions in simple ways.

%{\color{red} \sec{Acknowledgement}

%We thank a reviewer for bringing to our attention \cite{tor23a,tor23b}.}

\newpage

\sec{Appendix A: Scalar transformations}

In this appendix, we derive the one-dimensional Lorentz transformation from first principles. The principle of relativity states that the laws of physics are identical in all inertial frames. Consider two observers, one in the laboratory frame (LF) and the other in a moving frame (MF), which moves with (dimensionless) velocity $\bt$ relative to the LF. (We use the same units for $t$ and $x$, so $t$ is really $ct$ or $x$ is really $x/c$.) In the MF, the LF appears to move with velocity $-\bt$. (The observers agree on the magnitude of their relative velocity, but disagree  on its sign.) Let $(t, x)$ and $(t', x')$ be the time and distance coordinates in the LF and MF, respectively. Suppose that the origins of the coordinate systems coincide ($x' = x = 0$ at $t' = t = 0$) and light is emitted by a source at the common origin. Then the principle of relativity asserts that both observes will see the edges of the light wave move to the left and right (backward and forward) with unit speed. Mathematically, the squared intervals satisfy the equations
\be (t')^2 - (x')^2 = t^2 - x^2 = 0. \label{a1} \ee
What relations between the LF and MF coordinates allow these equations to be satisfied?

Following Occam's razor, suppose that the LF and MF coordinates are related by the linear equations
\be t' = at + bx, \ \ x' = ct + dx, \label{a2} \ee
where $a$, $b$, $c$ and $d$ are parameters.
Then the principle of continuity requires that $a$, $d \rightarrow 1$, and $b$, $c \rightarrow 0$ as $\bt \rightarrow 0$. The squared interval
\be (t')^2 - (x')^2 = (a^2 - c^2)t^2 + 2(ab - cd)tx - (d^2 - b^2)x^2. \label{a3} \ee
Equations (\ref{a1}) and (\ref{a3}) impose three conditions on the parameters, the first and third of which are
\be a = (1 + c^2)^{1/2}, \ \ d = (1 + b^2)^{1/2}. \label{a4} \ee
Although the interval condition allows both positive and negative roots, the continuity condition requires the positive roots. The second condition can be rewritten as $b/d = c/a$, from which it follows that
\be b/(1 + b^2)^{1/2} = c/(1 + c^2)^{1/2}. \label{a5} \ee
Hence, $b = c$. Equations (\ref{a2}) can be rewritten in the simpler forms
\be t' = at + bx, \ \ x' = ax + bt, \label{a6} \ee
where $a = (1 + b^2)^{1/2}$, or $b = \pm(a^2 - 1)^{1/2}$. It is customary to write $a = \ga$ and $b = -u$, where $\ga$ and $u$ are the (dimensionless) energy and momentum associated with the MF. (They are the energy and momentum of a particle of unit mass, which is at rest in the MF.) The velocity $\bt = u/\ga = (\ga^2 - 1)^{1/2}/\ga$, so $\ga = 1/(1 - \bt^2)^{1/2}$.
Notice that it is the momentum version of the transformation, rather than the velocity version, which arises naturally.

The squared interval $t^2 - x^2$ is an important concept. If the interval is positive, it is termed time-like (causal), because the emitted light has already arrived. If the interval is zero, the light is just arriving, and if the interval is negative, it is termed space-like (acausal), because the light has not yet arrived.

The rest of this article is based on the canonical equations
\be t' = \ga t + u x, \ \ x' = \ga x + u t. \label{a7} \ee
Suppose that $u > 0$. Then, in the passive picture (which was used above), the LF is stationary and the MF is moving backward (to the left) relative to the LF. In the active picture, the MF is stationary and the LF is moving forward (to the right) relative to the MF. A particle of unit mass that is at rest in the LF appears to have momentum $u$ and energy $\ga$ in the MF. Both pictures are valid, because of the relativistic equivalence of frames. Alternatively, one can consider only the LF, and regard $(t, x)$ and $(t', x')$ as the coordinate vectors before and after the transformation, respectively. In this picture, the transformation is called a boost.

\newpage

\sec{Appendix B: Vector transformations}

In this appendix, we consider Lorentz transformations in time and two space dimensions. (Active and passive transformations were defined and discussed in App. A.) Let $t$, $x$ and $y$ be the components of the coordinate three-vector. Then, for a boost in the $x$ direction,
\be t' = \ga t + u x, \ \ x' = \ga x + u t, \ \ y' = y, \label{b1} \ee
where $u$ and $\ga = (1 + u^2)^{1/2}$ are the (dimensionless) momentum and energy parameters, respectively. Time dilation depends on $\ga$ and the parallel distance. In the parallel direction, distance is dilated, whereas in the perpendicular direction, distance is conserved. If light is emitted from the origin, then at time $t$ its wavefront is a circle of radius $r = t$. Let $\Si = t^2 - r^2$ be the squared interval, which is also called the light-cone variable. Then if $\Si > 0$, the emitted light has already arrived at the point $(t, \vr)$, whereas if $\Si = 0$ it is just arriving, and if $\Si < 0$, it has not yet arrived. Furthermore,
\be \Si' = [(t')^2 - (x')^2] - (y')^2 = (t^2 - x^2) - y^2 = \Si. \label{b2} \ee
Transformation (\ref{b1}) conserves the squared interval, as it was designed to do (App. A).
Although we do not discuss rotations in this appendix, they also conserve the squared interval (because they conserve the squared distance $r^2$).

It is convenient to rewrite the transformation equations in vector form and determine their consequences by using only vector operations (addition, subtraction, and dot and cross products).
Suppose that $\vu = u\vn$ points in an arbitrary direction in the $xy$ plane. Then the components of $\vr$ parallel and perpendicular to $\vn$ are $\vr_\pa = \vn(\vn\cdot\vr)$ and $\vr_\perp = \vr - \vn(\vn\cdot\vr)$, respectively. In vector notation, the transformation equations are
\ba t' &= &\ga t + ur_\pa \nonumber \\
&= &\ga t + \vu\cdot\vr, \label{b3} \\
\vr\,' &= &\ga\vr_\pa + \vu t + \vr_\pe \nonumber \\
&= &[1 + (\ga - 1)\vn\vn\cdot] \vr + \vu t \nonumber \\
&= &[1 + \ep\vu\vu\cdot] \vr + \vu t, \label{b4} \ea
where $\ep = (\ga - 1)/(\ga^2 - 1) = 1/(\ga + 1)$. Let $\th = \vu\cdot\vr$. Then it is easy to verify that
\ba (t')^2 &= &\ga^2t^2 + 2\ga t\th + \th^2, \label{b5} \\
(r')^2 &= &[\vr + (t + \ep\th)\vu\,]^2 \nonumber \\
&= &r^2 + 2(t + \ep\th)\th + (t + \ep\th)^2u^2 \nonumber \\
&= &r^2 + 2(1 + \ep u^2)t\th + \ep(2 + \ep u^2)\th^2 + u^2t^2 \nonumber \\
&= &r^2 + 2\ga t\th + \th^2 + (\ga^2 - 1)t^2. \label{b6} \ea
Hence, $(t')^2 - (r')^2 = t^2 - r^2$, as stated above for the component equations (\ref{b1}). Notice that this result is a consequence of the transformation, not $t$ and $\vr$, which are arbitrary.

There are two ways to identify the significance of transformations (\ref{b3}) and (\ref{b4}). First, suppose that the increments $dt \neq 0$ and $d\vr = 0$. Then $dt' = \ga dt$ and $d\vr\,' = \vu dt$. Hence, the velocity $\vv\,' = d\vr\,'/dt' = \vu/\ga = \vv$. This result shows that the transformation is a boost (positive impulse). The inverse of a boost is a brake (negative impulse of the same magnitude).

Second, not only do Eqs. (\ref{b3}) and (\ref{b4}) apply to the three-vector $(t, \vr\,)$, they also apply to any other three-vector, such as the energy-momentum three-vector of a particle of unit mass $(\ga_p, \vu_p)$, whose components satisfy the identity $\ga_p^2 - u_p^2 = 1$, or the wave three-vector $(\om, \vk\,)$.
For a particle at rest, $\ga_p = 1$ and $\vu_p = 0$. According to Eqs. (\ref{b3}) and (\ref{b4}),
\be \ga_p' = \ga, \ \ \vu_p' = \vu. \label{b7} \ee
The transformed three-vector of the particle equals the three-vector that specified the transformation. Notice that $(\ga_p')^2 - (u_p')^2 = \ga^2 - u^2 = 1$. The difference between the squared energy and momentum is another Lorentz invariant, as is the difference betwen the squared frequency and wavevector.

Let $T = (t, \vr\,)$ be a coordinate three-vector. Then Eqs. (\ref{b3}) and (\ref{b4}) can be written in the tensor (matrix-like) form $T' = LT$, where the boost operator
\be L(\vu) = \left[\begin{array}{c|c} \ga & \vu\cdot \\ \hline \vu & 1 + \ep\vu\vu\cdot \end{array}\right]. \label{b8} \ee
The lines on the right side of Eq. (\ref{b8}) divide the operator into blocks. The top-left entry acts on a scalar to produce another scalar, the top-right entry acts on a vector to produce a scalar, the bottom-left entry acts on a scalar to produce a vector and the bottom-right entry acts on a vector to produce another vector. Notice that a boost operator is symmetric and is specified by two parameters, $u_x$ and $u_y$, or $\ga$ and $\th = \tan^{-1}(u_y/u_x)$. The opposite of a boost is a brake, so
\be L^{-1}(\vu) = L(-\vu) = \left[\begin{array}{c|c} \ga & -\vu\cdot \\ \hline -\vu & 1 + \ep\vu\vu\cdot \end{array}\right]. \label{b9} \ee
Notice that the top-left and bottom-right entries of $L$ are even functions of $\vu$, which do not change sign upon inversion. The matrix-like equation $L(-\vu)L(\vu) = I$ is equivalent to the constituent equations
\ba \ga^2 - u^2 &= &1, \label{b10} \\
\ga\vu\cdot - \vu\cdot(1 + \ep\vu\vu\cdot) &= &0, \label{b11} \\
- \ga\vu + (1 + \ep\vu\vu\cdot)\vu &= &0, \label{b12} \\
-\vu\vu\cdot + (1 + \ep\vu\vu\cdot)(1 + \ep\vu\vu\cdot) &= &1. \label{b13} \ea
It is easy to verify that Eqs. (\ref{b10}) -- (\ref{b13}) are satisfied, so operator (\ref{b9}) is the inverse of operator (\ref{b8}), as stated above.
For reference, $\ep\vu\vu = \de\vn\vn$, where $\de = \ga - 1$.

The three-vector of a particle at rest was transformed in Eq. (\ref{b7}).  Now consider the transformation of the three-vector of a particle that is already moving.
It follows from Eqs. (\ref{b3}) and (\ref{b4}), with $(\ga, \vu) = (\ga_2, \vu_2)$ and $(t, \vr) = (\ga_1, \vu_1)$,~that
\ba \ga_{21} &= &\ga_2\ga_1 + \vu_2\cdot\vu_1, \label{b21} \\
\vu_{21} &= &\vu_2\ga_1 + [1 + \ep_2\vu_2\vu_2\cdot] \vu_1 \nonumber \\
&= &\vu_1 + (\ga_1 + \ep_2\vu_2\cdot\vu_1)\vu_2. \label{b22} \ea
%
%Notice that formula (\ref{b21}) is symmetric in the indices 1 and 2.
Equations (\ref{b21}) and (\ref{b22}) are equivalent to the component equations (\ref{3.4.11}) -- (\ref{3.4.13}).
Let $\th = \vu_2\cdot\vu_1$. Then, by squaring Eqs. (\ref{b21}) and (\ref{b22}), one finds that
\ba \ga_{21}^2 &= &(\ga_2\ga_1)^2 + 2\ga_2\ga_1\th + \th^2, \label{b23} \\
u_{21}^2 &= &u_1^2 + 2(\ga_1 + \ep_2\th)\th + (\ga_1 + \ep_2\th)^2u_2^2. \label{b24} \ea
Notice the similarities between Eqs. (\ref{b23}) and (\ref{b24}), and Eqs. (\ref{b5}) and (\ref{b6}), respectively. In Eq. (\ref{b24}), the powers of $\th^0$, $\th^1$ and $\th^2$ are
\ba u_1^2 + \ga_1^2u_2^2 &= &\ga_1^2 - 1 + \ga_1^2(\ga_2^2 - 1) \ = \ (\ga_2\ga_1)^2 - 1, \label{b25} \\
2\ga_1(1 + \ep_2u_2^2) &= &2\ga_1(1 + \ga_2 - 1) \ = \ 2\ga_2\ga_1, \label{b26} \\
\ep_2(2 + \ep_2u_2^2) &= &\ep_2(2 + \ga_2 - 1) \ = \ 1, \label{b27} \ea
respectively.
By combining Eqs. (\ref{b23}) -- (\ref{b27}), one finds that $\ga_{21}^2 - u_{21}^2 = 1$. Thus, $(\ga_{21}, \vu_{21})$ is also an energy-momentum three-vector. This result was inevitable, because the second boost preserves the difference $\ga_1^2 - u_1^2 = 1$.

Two special cases are important. If the momenta $\vu_1$ and $\vu_2$ are parallel, then
\be \ga_{21} = \ga_2\ga_1 + u_2u_1, \ \ \vu_{21} = \ga_2\vu_1 + \ga_1\vu_2. \label{b28} \ee
Notice that formulas (\ref{b28}) are symmetric in the indices 1 and 2: For parallel boosts, the order of the boosts does not matter (Sec. 2). In contrast, if the momenta $\vu_1$ and $\vu_2$ are perpendicular, then
\be \ga_{21} = \ga_2\ga_1, \ \ \vu_{21} = \vu_1 + \ga_1\vu_2. \label{b29} \ee
The second of formulas (\ref{b29}) is not symmetric in the indices 1 and 2, so the boost order does matter (Secs. 3.4 and 3.9). Notice that $\vu_{21}$ is dominated by $\vu_2$, which is the boost momentum. This statement is generally true [Eq. (\ref{b22})], apart from exceptional cases in which $\vu_1$ and $\vu_2$ are nearly antiparallel. Also, notice that if $\vu_2 = -\vu_1$, then $\ga_{21} = 1$ and $\vu_{21} = 0$: The opposite of a boost is a brake of the same magnitude, as stated above.

If we had used $(\ga_1, \vu_1)$ to transform $(\ga_2, \vu_2)$, we would have found that
\ba \ga_{12} &= &\ga_1\ga_2 + \vu_1\cdot\vu_2, \label{b30} \\
\vu_{12} &= &\vu_2 + (\ga_2 + \ep_1\vu_1\cdot\vu_2)\vu_1, \label{b31} \ea
where $\ga_{12}^2 - u_{12}^2 = 1$. Although $\vu_{12}$ and $\vu_{21}$ are different vectors (and the formulas for them in terms of $\vu_1$ and $\vu_2$ are complicated), their lengths are equal because their associated energies are equal ($\ga_{12} = \ga_{21} = \ga$). Only their directions are different, which means that one vector is a rotated version of the other (in the plane defined by $\vu_1$ and $\vu_2$).
By combining Eqs. (\ref{b22}) and (\ref{b31}), one finds that
\be \vu_{12}\times\vu_{21} = (b_{12}b_{21} - 1)\vu_1\times\vu_2, \label{b32} \ee
where $b_{12} = \ga_2 + \ep_1\vu_1\cdot\vu_2$ and $b_{21} = \ga_1 + \ep_2\vu_1\cdot\vu_2$. Notice that the right side of Eq. (\ref{b32}) is proportional to the sine of the angle between $\vu_1$ and $\vu_2$, wheras the left side is proportional to the sine of the angle between $\vu_{12}$ and $\vu_{21}$. If the former vectors are not parallel, then neither are the latter vectors. We will return to this result later.

Although we chose to write Lorentz transformations in terms of the energy $\ga$ and momentum $\vu$, many authors write them in terms of the energy and velocity $\vv = \vu/\ga$. In these terms,
\ba \ga_{21} &= &\ga_2\ga_1(1 + \vv_2\cdot\vv_1), \label{b36} \\
\vu_{21} %&= &\ga_2\ga_1[(1 + \ga_2\ep_2\vv_2\cdot\vv_1)\vv_2 + \vv_1/\ga_2] \nonumber \\
%&= &(\ga_2\ga_1)[\vv_2 + (1 - 1/\ga_2 + 1/\ga_2)\vv_\pa + \vv_\pe/\ga_2] \nonumber \\
%&= &(\ga_2\ga_1)(\vv_2 + \vv_\pa + \vv_\pe/\ga_2), \label{b37}
&= &\ga_2\ga_1\vv_2 + \ga_1[\vv_\pa + \vv_\pe + (\ga_2 - 1)\vv_\pa] \nonumber \\
&= &(\ga_2\ga_1)(\vv_2 + \vv_\pa + \vv_\pe/\ga_2), \label{b37} \ea
where $\vv_\pa$ and $\vv_\pe$ are the components of $\vv_1$ parallel and perpendicular to $\vv_2$, respectively. By dividing the right side of Eq. (\ref{b37}) by the right side of Eq. (\ref{b36}), one obtains the velocity addition formula \cite{ein05}
\be \vv_{21} = {\vv_2 + \vv_\pa + \vv_\pe/\ga_2 \over 1 + \vv_2\cdot\vv_1}. \label{b38} \ee
Notice that $\vv_\pa$ and $\vv_\pe$ are transformed in different ways. Notice also that formula (\ref{b38}) reduces to formula (\ref{2.8}) in the limit $\vv_\pe \rightarrow 0$, as it should do.
It is easy to verify that
\ba (\vv_2 + \vv_\pa + \vv_\pe/\ga_2)^2 &= &(\vv_2 + \vv_\pa)^2 + v_\pe^2(1 - v_2^2) \nonumber \\
&= &(\vv_2 + \vv_1)^2 - (\vv_2\times\vv_1)^2. \label{b39} \ea
By combining Eqs. (\ref{b38}) and (\ref{b39}), one obtains the squared velocity formula \cite{ein05}
\be v_{21}^2 = {(\vv_2 + \vv_1)^2 - (\vv_2\times\vv_1)^2 \over (1 + \vv_2\cdot\vv_1)^2}. \label{b40} \ee

Now consider two successive transformations, and let $(\ga_1, \vu_1)$ and $(\ga_2, \vu_2)$ be three-vectors. Then, by applying Eqs. (\ref{b3}) and (\ref{b4}) twice, one finds that
\ba t'' &= &\ga_2(\ga_1t + \vu_1\cdot\vr\,) + \vu_2\cdot[\vu_1t + (1 + \ep_1\vu_1\vu_1\cdot)\vr\,] \nonumber \\
&= &(\ga_2\ga_1 + \vu_2\cdot\vu_1)t + [\vu_2\cdot + (\ga_2 + \ep_1\vu_2\cdot\vu_1)\vu_1\cdot]\vr, \label{b41} \\
\vr\,'' &= &\vu_2(\ga_1t + \vu_1\cdot\vr\,) + (1 + \ep_2\vu_2\vu_2\cdot)[\vu_1t + (1 + \ep_1\vu_1\vu_1\cdot)\vr\,] \nonumber \\
&= &[\vu_1 + (\ga_1 + \ep_2\vu_2\cdot\vu_1)\vu_2]t
+ [1 + \ep_2\vu_2\vu_2\cdot +\ \ep_1\vu_1\vu_1\cdot \nonumber \\
&&+\ (1 + \ep_2\ep_1\vu_2\cdot\vu_1)\vu_2\vu_1\cdot]\vr. \label{b42} \ea
Equations (\ref{b41}) and (\ref{b42}) can be written in the matrix-like form $T'' = L_2L_1T$, where the transformation operator
\be L_2(\vu_2)L_1(\vu_1) = \left[\begin{array}{c|c} \ga & \vu_{12}\cdot \\ \hline \vu_{21} & \st_{21}\cdot \end{array}\right]. \label{b43} \ee
The consituents of this operator are
\ba \ga &= &\ga_2\ga_1 + \vu_2\cdot\vu_1, \label{b44} \\
\vu_{12} &= &\vu_2 + (\ga_2 + \ep_1\vu_2\cdot\vu_1)\vu_1, \label{b45} \\
\vu_{21} &= &\vu_1 + (\ga_1 + \ep_2\vu_2\cdot\vu_1)\vu_2, \label{b46} \\
\st_{21} &= &1 + \ep_2\vu_2\vu_2 + \ep_1\vu_1\vu_1 + (1 + \ep_2\ep_1\vu_2\cdot\vu_1)\vu_2\vu_1. \label{b47} \ea
Equations (\ref{b44}) -- (\ref{b47}) are equivalent to the component equations (\ref{3.4.2}) -- (\ref{3.4.10}). [One can prove this statement by collecting the terms in the earlier equations that are proportional to the third and fourth powers of $c_i$ or $s_i$, and rewriting them in terms of $c_{21} = c_2c_1 + s_2s_1$, and the first and second powers of $c_i$ or $s_i$.]
%For reference,  $\ep_i\vu_i\vu_i = \de_i\vn_i\vn_i$.

If the operator $L_2L_1$ were to act on the three-vector $(1, 0)$, it would produce the three-vector $(\ga_{21}, \vu_{21})$, and if the operator $L_1L_2$ were to act on the three-vector $(1, 0)$, it would produce the three-vector $(\ga_{12}, \vu_{12})$, where $\ga_{12} = \ga_{21} = \ga$. This observation establishes the physical significances of the scalar in Eq. (\ref{b44}), and the vectors in Eqs. (\ref{b45}) and (\ref{b46}). However, the physical significance of the dyadic in Eq. (\ref{b47}) remains a mystery. Notice that this dyadic depends quadratically on $\vu_2$ and $\vu_1$, and is almost symmetric: The only term that depends on the order of the subscripts (boosts) is the last one, which is proportional to $\vu_2\vu_1$. Because of this asymmetry, the composition of two boosts is not another boost (unless $\vu_2$ is parallel to $\vu_1$).

In Sec. 3.5, it was shown that the composition of two boosts is a boost followed (or preceded) by a rotation.
Furthermore, the rotation (Wigner) angle is the angle between the row vector $\vu_{12}$ and the column vector $\vu_{21}$. It follows from Eq. (\ref{b32}) that
\be \sin(\th_w) = {(b_{12}b_{21} - 1)u_2u_1s_{21} \over \ga^2 - 1}, \label{b48} \ee
where $\ga$ was defined in Eq. (\ref{b44}), $s_{21} = \sin(\th_{21})$ and $\th_{21}$ is the angle between $\vu_1$ and $\vu_2$. Although Eq. (\ref{b48}) looks complicated, the numerator and denominator are both divisible by $\ga - 1$. The numerator is proportional to
\ba &&(b_{12}b_{21} - 1)u_2u_1 \nonumber \\
&= &(\ga_2u_1 + u_2\de_1c_{21})(u_2\ga_1 + \de_2u_1c_{21}) - u_2u_1 \nonumber \\
&= &(\ga_2\ga_1 - 1)u_2u_1 + (\ga_2\de_2u_1^2 + u_2^2\ga_1\de_1)c_{21} + \de_2\de_1u_2u_1c_{21}^2, \label{b49} \ea
where $c_{21} = \cos(\th_{21})$.
The first term in Eq. (\ref{b49}) is the product of $\ga_2\ga_1 - 1$ and $u_2u_1$, and the last term is the product of $u_2u_1c_{21}$ and $\de_2\de_1c_{21}$, so consider the ansatz
\ba &&(\ga_2\ga_1 - 1 + u_2u_1c_{21})(u_2u_1 + \de_2\de_1c_{21}) \nonumber \\
&= &(\ga_2\ga_1 - 1)u_2u_1 + [(\ga_2\ga_1 - 1)\de_2\de_1 + u_2^2u_1^2]c_{21} + \de_2\de_1u_2u_1c_{21}^2. \label{b50} \ea
In Eqs. (\ref{b49}) and (\ref{b50}), the coefficients of $c_{21}$ both equal $(2\ga_2\ga_1 + \ga_2 + \ga_1)\de_2\de_1$, so the latter equation is the required factorization of the former. By combining the preceding results, one finds that
\be \sin(\th_w) = {(u_2u_1 + \de_2\de_1c_{21})s_{21} \over \ga_2\ga_1 + 1 + u_2u_1c_{21}}. \label{b51} \ee
Equation (\ref{b51}) is consistent with Eq. (\ref{3.9.26}) and the sine formula stated after Eq. (\ref{3.5.13}).

The inverse of two boosts is two brakes, done in reverse order. (Undo boost 2, then undo boost 1.) According to Eq. (\ref{b9}), the brake associated with the boost $L(\vu)$ is $L(-\vu)$. Hence, the inverse transformation operator
\be [L_2(\vu_2)L_1(\vu_1)]^{-1} = L_1(-\vu_1)L_2(-\vu_2) = \left[\begin{array}{c|c} \ga & -\vu_{21}\cdot \\ \hline -\vu_{12} & \st_{12}\cdot \end{array}\right], \label{b52} \ee
where $\ga$, $\vu_{12}$ and $\vu_{21}$ were defined above, and one obtains the formula for $\st_{12}$ from the formula for $\st_{21}$ by interchanging the subscripts 1 and 2. Thus, the inverse of the operator is its transpose, with the signs of $\vu_{12}$ and $\vu_{21}$ negated. This result is consistent with Eq. (\ref{3.1.2}).

Although the main results of this section [Eqs. (\ref{b3}), (\ref{b4}) and (\ref{b8}), Eqs. (\ref{b21}) and (\ref{b22}), and Eqs. (\ref{b41}) -- (\ref{b43})] were derived for two space dimensions, they are also valid for three dimensions, because they are based on the laws of vector algebra. The main difference between the two cases is that in two dimensions, the third (successive) transformation is coplanar (in the plane defined by the momentum vectors of the first two transformations), whereas in three dimensions, the third transformation need not be coplanar. Vector transformations, including rotations, are discussed in App. C and \cite{cus67}. 

\newpage

\sec{Appendix C: Alternative product tensor}

In App. B, we showed that the product operator for two boosts, of momenta $\vu_1$ and $\vu_2$, can be written in the tensor form
\be L_{21} = \left[\begin{array}{c|c} \ga & \vu_{12}\cdot \\ \hline \vu_{21} & \st_{21} \end{array}\right], \label{c1} \ee
where $|u_{12}|^2 = |u_{21}|^2 = \ga^2 - 1$.
The consituents of this tensor were specified in Eqs. (\ref{b44}) -- (\ref{b47}).
The combination of two nonparallel boosts is not a boost, because the product tensor is asymmetric. Rather, it is a boost followed by a rotation [Eq. (\ref{3.5.11}) and the text after Eq. (\ref{3.5.14})].
Let $\ot_{21}$ be the rotation operator that converts $\vu_{12}$ to $\vu_{21}$. Then the product operator
\ba L_{21} &= &\left[\begin{array}{c|c} 1 & 0 \\ \hline 0 & \ot_{21} \end{array}\right]
\left[\begin{array}{c|c} \ga & \vu_{12}\cdot \\ \hline \vu_{12} & 1 + \ep\vu_{12}\vu_{12}\cdot \end{array}\right] \nonumber \\
&= &\left[\begin{array}{c|c} \ga & \vu_{12}\cdot \\ \hline \vu_{21} & \ot_{21} +\ \ep\vu_{21}\vu_{12}\cdot \end{array}\right], \label{c2} \ea
where $\ep = 1/(\ga + 1)$. By comparing Eqs. (\ref{c1}) and (\ref{c2}), one finds that the spatial operator
\be \st_{21} = \ot_{21} + \ep\vu_{21}\vu_{12}\cdot. \label{c3} \ee
In Eq. (\ref{b47}), $\st_{21}$ was specified in terms of $\vu_1$ and $\vu_2$, whereas in Eq. (\ref{c3}), it is specified in terms of $\vu_{12}$ and $\vu_{21}$. The dyadic is specified explicitly, whereas the rotation operator is specified implicitly.

Let $\vn_1$ and $\vn_2$ be arbitrary unit vectors and let $\th_{21}$ be the angle between them. Then
\be \ot_{21} = (\vn_1\cdot\vn_2) + (\vn_1\times\vn_2)\times, \label{c4} \ee
where the dot product has magnitude $c_{21} = \cos(\th_{21})$ and the cross product has magnitude $s_{21} = \sin(\th_{21})$. It is easy to verify that $\ot_{21}\vn_1 = \vn_2$.
In the context of transformation (\ref{c2}),
\be \ot_{21} = {\vu_{12}\cdot\vu_{21} + (\vu_{12}\times\vu_{21})\times \over \ga^2 - 1}. \label{c5} \ee
Taken together, Eqs. (\ref{c3}) and (\ref{c5}) specify the spatial operator explicitly.
The angle between $\vu_{12}$ and $\vu_{21}$ is called the Wigner angle ($\th_w$).
It remains to be shown that the alternative formula (\ref{c3}) is equivalent to the standard formula (\ref{b47}).

The cross-product term in Eq. (\ref{c5}) can be rewritten as an antisymmetric antidiagonal matrix, the magnitude of whose components is the sine of the Wigner angle. By combining Eqs. (\ref{b32}), (\ref{b48}) and (\ref{b50}), one finds that
\be {(\vu_{12}\times\vu_{21})\times \over \ga^2 - 1} = {1 \over \ga + 1} \left[\begin{array}{cc} 0 & -(u_2u_1 + \de_2\de_1c)s \\ (u_2u_1 + \de_2\de_1c)s & 0 \end{array}\right], \label{c11} \ee
where $c$ and $s$ are abbreviations for $c_{21}$ and $s_{21}$, respectively.
For reference, the derivation of Eq. (\ref{c11}) utilized the fact that the magnitude of the cross product is divisible by $\ga - 1$.

The dot-product term in Eq. (\ref{c5}) can be rewritten as a diagonal matrix, the magnitude of whose components is the cosine of the Wigner angle. The dot product
\ba \vu_{12}\cdot\vu_{21} &= &(\vu_2 + b_{12}\vu_1)\cdot(\vu_1 + b_{21}\vu_2) \nonumber \\
&= &b_{21}u_2^2 + b_{12}u_1^2 + (b_{12}b_{21} + 1)\vu_2\cdot\vu_1, \label{c12} \ea
where $b_{12} = \ga_2 + \ep_1\vu_2\cdot\vu_1$ and $b_{21} = \ga_1 + \ep_2\vu_2\cdot\vu_1$. It is easy to verify that
\ba b_{21}u_2^2 + b_{12}u_1^2 &= &(u_2\ga_1 + \de_2u_1c)u_2 + (\ga_2u_1 + u_2\de_1c)u_1, \label{c13} \\
(b_{12}b_{21} + 1)u_2u_1 &= &(\ga_2u_1 + u_2\de_1c)(u_2\ga_1 + \de_2u_1c) + u_2u_1. \label{c14} \ea
By combining Eqs. (\ref{c12}) -- (\ref{c14}), one obtains a formula for the dot product, in which the coefficients of the powers of $c$ are
\ba &c^0: &\ga_2(\ga_1^2 - 1) + (\ga_2^2 - 1)\ga_1 \nonumber \\
&&=\ (\ga_2\ga_1 - 1)(\ga_2 + \ga_1), \label{c15} \\
&c^1: &(\de_2 + \de_1)u_2u_1 + (\ga_2\ga_1 + 1)u_2u_1 \nonumber \\
&&=\ (\ga_2\ga_1 - 1 + \ga_2 + \ga_1)u_2u_1, \label{c16} \\
&c^2: &\ga_2\de_2u_1^2 + u_2^2\ga_1\de_1, \label{c17}  \\
&c^3: &\de_2\de_1u_2u_1. \label{c18} \ea
The magnitude of the cross product was divisible by $\ga - 1$, so it is natural to ask if the dot product is divisible by the same factor. Consider the ansatz
\be (\ga_2\ga_1 - 1 + u_2u_1c)[\ga_2 + \ga_1 + (u_2u_1 + \de_2\de_1c)c]. \label{c19} \ee
It is easy to verify that the coefficients of $c^0$, $c^1$ and $c^3$ in ansatz (\ref{c19}) are specified by Eqs. (\ref{c15}), (\ref{c16}) and (\ref{c18}), respectively. The coefficient of $c^2$ is
\be (\ga_2\ga_1 - 1)\de_2\de_1 + u_2^2u_1^2. \label{c20}\ee
Although coefficients (\ref{c17}) and (\ref{c20}) look different, they both equal $(2\ga_2\ga_1 + \ga_2 + \ga_1)\de_2\de_1$, so Eq. (\ref{c19}) is the required factorization of Eq. (\ref{c12}). Consequently,
\be {\vu_{12}\cdot\vu_{21} \over \ga^2 - 1} = {1 \over \ga + 1} \left[\begin{array}{cc} \ga_2 + \ga_1 + (u_2u_1 + \de_2\de_1c)c & 0 \\ 0 & \ga_2 + \ga_1 + (u_2u_1 + \de_2\de_1c)c \end{array}\right]. \label{c21} \ee
By combining Eqs. (\ref{c11}) and (\ref{c21}), one obtains a rotation matrix ($\Om_{21}$)
%
%\be \Om_{21} = {1 \over \ga + 1} \left[\begin{array}{cc} \ga_2 + \ga_1 + (u_2u_1 + \de_2\de_1c)c & -(u_2u_1 + \de_2\de_1c)s \\ (u_2u_1 + \de_2\de_1c)s & \ga_2 + \ga_1 + (u_2u_1 + \de_2\de_1c)c \end{array}\right], \label{c23}  \ee
%
whose diagonal and nondiagonal components are $c_w$ and $\mp s_w$, respectively.
[See the text between Eqs. (\ref{3.5.13}) and (\ref{3.5.14}).]

Now consider the dyadic $\ep\vu_{21}\vu_{12}$. It follows from Eqs. (\ref{b45}) and (\ref{b46}) that
\ba \vu_{21}\vu_{12} &= &(\vu_1 + b_{21}\vu_2)(\vu_2 + b_{12}\vu_1) \nonumber \\
&= &\vu_1\vu_2 + b_{12}\vu_1\vu_1 + b_{21}\vu_2\vu_2 + b_{21}b_{12}\vu_2\vu_1, \label{c31} \ea
where $b_{12} = \ga_2 + \ep_1\vu_2\cdot\vu_1$ and $b_{21}= \ga_1 + \ep_2\vu_2\cdot\vu_1$.
One can rewrite the constituent dyadics in the matrix forms
\ba \vu_1\vu_2 &= &u_2u_1 \left[\begin{array}{cc} c_1c_2 & c_1s_2 \\ s_1c_2 & s_1s_2 \end{array}\right], \label{c32} \\
b_{12}\vu_1\vu_1 &= &(\ga_2u_1 + u_2\de_1c)u_1 \left[\begin{array}{cc} c_1^2 & c_1s_1 \\ s_1c_1 & s_1^2 \end{array}\right], \label{c33} \\
b_{21}\vu_2\vu_2 &= &(u_2\ga_1 + \de_2u_1c)u_2 \left[\begin{array}{cc} c_2^2 & c_2s_2 \\ s_2c_2 & s_2^2 \end{array}\right], \label{c34} \\
b_{21}b_{12}\vu_2\vu_1 &= &(\ga_2u_1 + u_2\de_1c)(u_2\ga_1 + \de_2u_1c)
\left[\begin{array}{cc} c_2c_1 & c_2s_1 \\ s_2c_1 & s_2s_1 \end{array}\right], \label{c35} \ea
where the coefficients
\ba \ga_2u_1^2 + u_2u_1\de_1c &= &[\ga_2(\ga_1 + 1) + u_2u_1c]\de_1 \nonumber \\
&= &(\ga + \ga_2)\de_1, \label{c36} \\
 u_2^2\ga_1 + u_2u_1\de_2c &= &[(\ga_2 + 1)\ga_1 + u_2u_1c]\de_2 \nonumber \\
&= &(\ga + \ga_1)\de_2. \label{c37} \ea
Notice that $\ga + \ga_i = \ga + 1 + \de_i$.

Let $D_{21} = [d_{ij}]$ be the matrix form of the dyadic $\vu_{21}\vu_{12}/(\ga + 1)$. %(This matrix is not diagonal.)
Two components of this matrix are
\ba (\ga + 1)d_{11} &= &u_2u_1c_2c_1 + (\ga + \ga_1)\de_2c_2^2 + (\ga + \ga_2)\de_1c_1^2 \nonumber \\
&&+\ (\ga_2u_1 + u_2\de_1c)(u_2\ga_1 + \de_2u_1c)c_2c_1, \label{c41} \\
(\ga + 1)d_{21} &= &u_2u_1s_1c_2 + (\ga + \ga_1)\de_2s_2c_2 + (\ga + \ga_2)\de_1s_1c_1 \nonumber \\
&&+\ (\ga_2u_1 + u_2\de_1c)(u_2\ga_1 + \de_2u_1c)s_2c_1. \label{c42} \ea
%
%I did not notice useful simplifications of the first and second lines of Eqs. (\ref{d8}) and (\ref{d9}), so I added to the third lines the associated components of matrix (\ref{c23}), to form the matrix analog of the whole tensor $\st_{21}$. To make sense of formulas (\ref{d8}) and (\ref{d9}), they must be simplified.
For reference,
\ba (\ga + 1)(u_2u_1 + \de_2\de_1c) &= &(\ga_2\ga_1 + 1)u_2u_1 + [(\ga_2\ga_1 + 1)\de_2\de_1 + u_2^2u_1^2]c \nonumber \\
&&+\ u_2u_1\de_2\de_1c^2, \label{c43} \ea
where the coefficient of the $c$ term equals $(2\ga_2\ga_1 + \ga_2 + \ga_1 + 2)\de_2\de_1$.

By reordering the terms in Eq. (\ref{c41}), one finds that
\ba (\ga + 1)d_{11} &= &(\ga + 1)\de_2c_2^2 + (\ga + 1)\de_1c_1^2 + \de_2\de_1(c_2^2 + c_1^2) \nonumber \\
&&+\ [(\ga_2u_1 + u_2\de_1c)(u_2\ga_1 + \de_2u_1c) + u_2u_1]c_2c_1. \label{c44} \ea
The term in square brackets equals
\be (\ga_2\ga_1 + 1)u_2u_1 + (\ga_2\de_2u_1^2 + u_2^2\ga_1\de_1)c + u_2u_1\de_2\de_1c^2, \label{c45} \ee
where the coefficient of the $c$ term equals $(2\ga_2\ga_1 + \ga_2 + \ga_1)\de_2\de_1$. This coefficient differs from the coefficient in Eq. (\ref{c43}) by $-2\de_2\de_1c$. Hence,
\ba (\ga + 1)d_{11} &= &(\ga + 1)\de_2c_2^2 + (\ga + 1)\de_1c_1^2 \nonumber \\
&&+\ (\ga + 1)(u_2u_1 + \de_2\de_1c)c_2c_1 \nonumber \\
&&+\ \de_2\de_1(c_2^2 + c_1^2) - 2\de_2\de_1cc_2c_1. \label{c46} \ea
In the third line of Eq. (\ref{c46}), the coefficient of $\de_2\de_1$ term is
\ba &&c_2^2(c_1^2 + s_1^2) + (c_2^2 + s_2^2)c_1^2 - 2(c_2c_1+ s_2s_1)c_2c_1 \nonumber \\
&= &c_2^2s_1^2 + s_2^2c_1^2 - 2s_2s_1c_2c_1 \nonumber \\
&= &(s_2c_1 - c_2s_1)^2. \label{c47} \ea
It follows from Eqs. (\ref{c46}) and (\ref{c47}) that
%
%\ba (\ga + 1)d_{11} &= &(\ga + 1)\de_2c_2^2 + (\ga + 1)\de_1c_1^2 \nonumber \\
%&&+\ (\ga + 1)(u_2u_1 + \de_2\de_1c)c_2c_1 + \de_2\de_1s^2. \label{d15} \ea
%
%The factors of $\ga + 1$ on the left and right sides of Eq. (\ref{d15}) do not cancel completely.
%
\ba d_{11} &= &\de_2c_2^2 + \de_1c_1^2 + (u_2u_1 + \de_2\de_1c)c_2c_1 \nonumber \\
&&+\ \de_2\de_1s^2/(\ga + 1). \label{c48} \ea

By reordering the terms in Eq. (\ref{c42}), one finds that
\ba (\ga + 1)d_{21} &= &(\ga + 1)\de_2s_2c_2 + (\ga + 1)\de_1s_1c_1 + \de_2\de_1(s_2c_2 + s_1c_1) \nonumber \\
&&+\ [(\ga_2u_1 + u_2\de_1c)(u_2\ga_1 + \de_2u_1c) + u_2u_1]s_2c_1 \nonumber \\
&&+\ u_2u_1(s_1c_2 - s_2c_1) \nonumber \\
&= &(\ga + 1)\de_2s_2c_2 + (\ga + 1)\de_1s_1c_1 \nonumber \\
&&+\ (\ga + 1)(u_2u_1 + \de_2\de_1c)s_2c_1 - u_2u_1(s_2c_1 -c_2s_1) \nonumber \\
&&+\ \de_2\de_1(s_2c_2 + s_1c_1) - 2\de_2\de_1cs_2c_1. \label{c49} \ea
In the last line of Eq. (\ref{c49}), the coefficient of $\de_2\de_1$ is
\ba &&s_2c_2(c_1^2 + s_1^2) + (c_2^2 + s_2^2)s_1c_1 - 2(c_2c_1 + s_2s_1)s_2c_1 \nonumber \\
&= &-\ s_2c_1(c_2c_1 + s_2s_1) + (c_2c_1 + s_2s_1)s_1c_2 \nonumber\\
&= &-(c_2c_1 + s_2s_1)(s_2c_1 - c_2s_1). \label{c50} \ea
It follows from Eqs. (\ref{c49}) and (\ref{c50}) that
%
%\ba (\ga + 1)d_{21} &= &(\ga + 1)\de_2s_2c_2 + (\ga + 1)\de_1s_1c_1 \nonumber \\
%&&+\ (\ga + 1)(u_2u_1 + \de_2\de_1c)s_2c_1 \nonumber \\
%&&-\ (u_2u_1 + \de_2\de_1c)s. \label{d18} \ea
%
%Once again, the coefficients of $\ga + 1$ on the left and right sides do not cancel completely.
%
\ba d_{21} &= &\de_2s_2c_2 + \de_1s_1c_1 + (u_2u_1 + \de_2\de_1c)s_2c_1 \nonumber \\
&&-\ (u_2u_1 + \de_2\de_1c)s/(\ga + 1). \label{c51} \ea

The analyses of the other terms are similar. The results are
%
%\ba (\ga + 1)d_{12} &= &(\ga + 1)\de_2c_2s_2 + (\ga + 1)\de_1c_1s_1 \nonumber \\
%&&+\ (\ga + 1)(u_2u_1 + \de_2\de_1c)c_2s_1 \nonumber \\
%&&+\ (u_2u_1 + \de_2\de_1c)s, \label{d19} \\
%(\ga + 1)d_{22} &= &(\ga + 1)\de_2s_2^2 + (\ga + 1)\de_1s_1^2 \nonumber \\
%&&+\ (\ga + 1)(u_2u_1 + \de_2\de_1c)s_2s_1 + \de_2\de_1s^2. \label{d20} \ea
%
%
\ba d_{12} &= &\de_2c_2s_2 + \de_1c_1s_1 + (u_2u_1 + \de_2\de_1c)c_2s_1 \nonumber \\
&&+\ (u_2u_1 + \de_2\de_1c)s/(\ga + 1), \label{c52} \\
d_{22} &= &\de_2s_2^2 + \de_1s_1^2 + (u_2u_1 + \de_2\de_1c)s_2s_1 \nonumber \\
&&+\ \de_2\de_1s^2/(\ga + 1). \label{c53} \ea
One can deduce Eqs. (\ref{c52}) and (\ref{c53}) from Eqs. (\ref{c48}) and (\ref{c51}), respectively, by interchanging $c_i$ and $s_i$, which changes the sign of $s$, but does not affect $c$.

Let $M_{21} = [m_{ij}]$ be the matrix form of the spatial tensor (\ref{b47}). Then, written explicitly,
\ba M &= &\left[\begin{array}{cc} 1 & 0 \\ 0 & 1 \end{array}\right]
+ \de_2\left[\begin{array}{cc} c_2^2 & c_2s_2 \\ s_2c_2 & s_2^2 \end{array}\right]
+ \de_1\left[\begin{array}{cc} c_1^2 & c_1s_1 \\ s_1c_1 & s_1^2 \end{array}\right ]\nonumber \\
&&+\ (u_2u_1 + \de_2\de_1c_{21})\left[\begin{array}{cc} c_2c_1 & c_2s_1 \\ s_2c_1 & s_2s_1 \end{array}\right].  \label{c56} \ea
By combining Eqs. (\ref{c21}) and (\ref{c48}), and Eqs. (\ref{c11}) and (\ref{c51}), one finds that
\ba  m_{11} &= &[\ga_2 + \ga_1 + (u_2u_1 + \de_2\de_1c)c]/(\ga + 1) \nonumber \\
&&+\ \de_2c_2^2 + \de_1c_1^2 + (u_2u_1 + \de_2\de_1c)c_2c_1 \nonumber \\
&&+\ \de_2\de_1s^2/(\ga + 1) \nonumber \\
&= &(\ga_2 + \ga_1 + \de_2\de_1+  u_2u_1c)/(\ga + 1) \nonumber \\
&&+\ \de_2c_2^2 + \de_1c_1^2 + (u_2u_1 + \de_2\de_1c)c_2c_1 \nonumber \\
&= & 1 + \de_2c_2^2 + \de_1c_1^2 + (u_2u_1 + \de_2\de_1c)c_2c_1, \label{c57} \\
m_{21} &= &(u_2u_1 + \de_2\de_1c)s/(\ga + 1) + \de_2s_2c_2 + \de_1s_1c_1 \nonumber \\
&&+\ (u_2u_1 + \de_2\de_1c)s_2c_1 - (u_2u_1 + \de_2\de_1c)s/(\ga + 1) \nonumber \\
&= &\de_2s_2c_2 + \de_1s_1c_1 + (u_2u_1 + \de_2\de_1c)s_2c_1, \label{c58} \ea
respectively.
Equations (\ref{c57}) and (\ref{c58}), and their counterparts for $m_{12}$ and $m_{22}$, are consistent with Eq. (\ref{c56}).
Thus, the the alternative formula (\ref{c3}) is equivalent to the standard formula (\ref{b47}).

Equation (\ref{c4}) can be rewritten in the form $\ot_{21} = c + s\vn\times$, where $\vn = (\vn_1\times\vn_2)/|\vn_1\times\vn_2|$ is the unit vector that defines the axis of rotation. In three dimensions, it remains true that $\ot_{21}\vn_1 = \vn_2$. Nonetheless, one should add to the rotation tensor the term $(1 - c)\vn\vn\cdot$, which allows the tensor to preserve the parallel components of the vectors on which it acts \cite{mck25c}.

\newpage

\sec{Appendix D: Fundamental transformations}

In a related paper \cite{mck25c}, we consider rotations in three space dimensions. Rotation matrices are specified by three free parameters. The standard set of parameters consists of two polar angles, which specify the rotation axis, and the rotation angle. We showed that every rotation matrix has the decomposition $R = R_zR_yR_x$, where $R_i$ represents a rotation about the $i$ axis through the angle $\th_i$. Because axial rotations are the building blocks of arbitrary rotations, they are termed fundamental. The three angles $\th_x$, $\th_y$ and $\th_z$ comprise an alternative set of parameters. It is natural to ask if a similar result exists for Lorentz matrices.

Let $B_x$ and $B_y$ represent boosts along the $x$ and $y$ axes, respectively, and $R_t$ represent a rotation about the $t$ axis (in the $xy$ plane). Each of these transformation matrices is the exponential of a generating matrix [Eqs. (\ref{3.8.2}) and (\ref{3.8.5})]. Because each matrix is a Lorentz matrix, so also is their product $R_tB_yB_x$.
In Sec. 3.9, we calculated the product matrix $B_yB_x$, which has the Schmidt-like decomposition $R_2B_xR_1^t$, where $R_i$ is a rotation matrix, so it is obvious that $R_tB_yB_x$ has the standard form $R_3B_xR_1^t$. What is not obvious is whether it is an arbitrary or a specific Lorentz matrix. Written explicitly, the product matrix
\ba R_tB_yB_x &= &\left[\begin{array}{ccc} 1 & 0 & 0 \\ 0 & c & -s \\ 0 & s & c \end{array}\right]
\left[\begin{array}{ccc} \ga_2\ga_1 & \ga_2u_1 & u_2 \\ u_1 & \ga_1 & 0 \\ u_2\ga_1 & u_2u_1 & \ga_2 \end{array}\right] \nonumber \\
&= &\left[\begin{array}{ccc} \ga_2\ga_1 & \ga_2u_1 & u_2 \\ cu_1 - su_2\ga_1 & c\ga_1 - su_2u_1 & -s\ga_2 \\ su_1 + cu_2\ga_1 & s\ga_1 + cu_2u_1 & c\ga_2 \end{array}\right], \label{d1} \ea
where $u_i$ is a momentum and $\ga_i = (1 + u_i^2)^{1/2}$ is an energy, and $c = \cos(\th)$ and $s = \sin(\th)$, where $\th$ is the rotation angle. Matrix (\ref{d1}) is specified by three free parameters: the momenta $u_1$ and $u_2$, and the angle $\th$.

Every Lorentz matrix $L$ can be written in the form of matrix (\ref{3.3.12}), which is also specified by three free parameters: the momentum $u$, and the angles $\th_1$ and $\th_2$. The energy $\ga = (1 + u^2)^{1/2}$ and the difference angle $\th_{21} = \th_2 - \th_1$.
By comparing Eqs. (\ref{3.3.12}) and (\ref{d1}), one finds that $u_2 = us_1$, which determines $\ga_2$, and $u_1 = uc_1/\ga_2$, which determines $\ga_1$. (Notice that the tangent $t_1 = u_2/\ga_2u_1$.) The only undetermined parameter in Eq. (\ref{d1}) is $\th$.

Henceforth, we will use the symbol $l_{ij}$ to denote the same component of $L$ and $R_tB_yB_x$. By dividing $l_{20}$ by $l_{10}$, one finds that
\be t_2 = {u_1t + u_2\ga_1 \over u_1 - u_2\ga_1t}, \ \ 
t = {u_1t_2 - u_2\ga_1 \over u_1 + u_2\ga_1t_2}, \label{d3} \ee
where $t = s/c$. By combining $l_{11}$ and $l_{22}$, one finds that
\be (\ga + 1)c_{21} = (\ga_2 + \ga_1)c - u_2u_1s, \label{d4} \ee
and by combining $l_{12}$ and $l_{21}$, one finds that
\be (\ga + 1)s_{21} = (\ga_2 + \ga_1)s + u_2u_1c. \label{d5} \ee
It follows from Eqs. (\ref{d4}) and (\ref{d5}) that
\be t_{21} = {(\ga_2 + \ga_1)t + u_2u_1 \over (\ga_2 + \ga_1) - u_2u_1t}, \ \ 
t = {(\ga_2 + \ga_1)t_{21} - u_2u_1 \over (\ga_2 + \ga_1) + u_2u_1t_{21}}. \label{d6} \ee
In principle, one can use the second of Eqs. (\ref{d3}) or the second of Eqs. (\ref{d6}) to specify $t$.

The aforementioned equations are consistent if and only if
\be (u_1t_2 - u_2\ga_1)[(\ga_2 + \ga_1) + u_2u_1t_{21}] = (u_1 + u_2\ga_1t_2)[(\ga_2 + \ga_1)t_{21} - u_2u_1]. \label{d12} \ee
By collecting terms on the right, one finds that the coefficient of $t_{21}$ is
\ba &&(u_1 + u_2\ga_1t_2)(\ga_2 +\ga_1) + (u_2\ga_1 - u_1t_2)u_2u_1 \nonumber \\
&= &[u_2(\ga_2\ga_1 + \ga_1^2) - (\ga_1^2 - 1)u_2]t_2 + u_1(\ga_2 + \ga_1) + (\ga_2^2 - 1)\ga_1u_1 \nonumber \\
&= &(\ga_2\ga_1 + 1)u_2t_2 + (\ga_2\ga_1 + 1)\ga_2u_1, \label{d13} \ea
and by collecting terms on the left, one finds that the coefficient of 1 is
\ba &&(u_1t_2 - u_2\ga_1)(\ga_2 + \ga_1) + (u_1 + u_2\ga_1t_2)u_2u_1 \nonumber \\
&= &[u_1(\ga_2 + \ga_1) + (\ga_2^2 - 1)\ga_1u_1]t_2 + u_2(\ga_1^2 - 1) - u_2(\ga_2\ga_1 + \ga_1^2) \nonumber \\
&= &(\ga_2\ga_1 + 1)\ga_2u_1t_2 - (\ga_2\ga_1 + 1)u_2. \label{d14} \ea
The two sides of Eq. (\ref{d12}) are equal if and only if
\be t_{21} = {\ga_2u_1t_2 - u_2 \over \ga_2u_1 + u_2t_2}
= {t_2 - u_2/\ga_2u_1 \over 1 + t_2u_2/\ga_2u_1}. \label{d15} \ee
Equation (\ref{d15}) is the standard formula for $t_{21}$ in terms of $t_2$ and $t_1 = u_2/\ga_2u_1$. Thus, the formulas for $t$ (and $\th$) are consistent.

The preceding analysis shows that one can choose the parameters $u_1$, $u_2$ and $\th$ in such a way that the product matrix $R_tB_yB_x$ equals an arbitrary Lorentz matrix $L$. Because axial boosts and a rotation are the building blocks of arbitrary transformations, they are termed fundamental.
This result is an example of normal ordering (Sec. 3.8): In the decomposition $L = R_tB_yB_x$, the powers of the generator $G_t$ appear to the left of (before) the powers of $G_y$, which appear before the powers of $G_x$.

\end{document}